\documentclass[12pt]{article}

\usepackage[fleqn]{amsmath}
\usepackage{amsfonts}
\usepackage{amssymb}
\usepackage{mathrsfs}
\usepackage{xcolor}

\usepackage{epsfig}
\usepackage{caption}
\usepackage{float}
\usepackage[title]{appendix}

\usepackage{hyperref}

\hypersetup{
    colorlinks,
    citecolor=blue,
    filecolor=blue,
	    linkcolor=blue,
    urlcolor=blue
}

\newcommand{\NN}{\mathbb{N}}
\newcommand{\RR}{\mathbb{R}}
\newcommand{\SP}{\mathbb{S}}
\newcommand{\ZZ}{\mathbb{Z}}

\newcommand{\darkred}[1]{\textcolor{black}{{#1}}}
\newcommand{\blue}[1]{\textcolor{black}{{#1}}}
\newcommand{\magenta}[1]{\textcolor{black}{{#1}}}

\newcommand{\beg}{\begin{equation}}
\newcommand{\en}{\end{equation}}

 \newcommand{\lam}{\lambda}

\newcommand{\eps}{\varepsilon}

\DeclareFontFamily{U}{mathx}{}
\DeclareFontShape{U}{mathx}{m}{n}{<-> mathx10}{}
\DeclareSymbolFont{mathx}{U}{mathx}{m}{n}

\DeclareMathAccent{\widecheck}{0}{mathx}{"71}

\begin{document}

\title{Bounds on $T_c$ in the Eliashberg theory of\\ Superconductivity. I: The $\gamma$-model \vspace{-0.5truecm}}
\author{M. K.-H. Kiessling,$^1\!$ B. L. Altshuler,$^2\!$  and E. A. Yuzbashyan$^3$\\ \small
          $^1$ Department of Mathematics,\\ \small
Rutgers, The State University of New Jersey,\\ \small
          110 Frelinghuysen Road, Piscataway, NJ 08854\\ \small
          $^2$ Department of Physics, \\ \small
Columbia University,\\ \small
             538 West 120th Street, New York, NY 10027\\ \small
             $^3$ Center of Materials Theory \\ \small Department of Physics and Astronomy, \\ \small
Rutgers, The State University of New Jersey,\\ \small
          136 Frelinghuysen Road, Piscataway, NJ 08854\vspace{-.3truecm}}

\date{Final version; April 04, 2025}
\maketitle

\vspace{-1truecm}
\abstract{\noindent
{Using the recent reformulation   for the Eliashberg theory of superconductivity}
in terms of a classical interacting Bloch spin chain model, rigorous upper and lower bounds on the critical temperature
$T_c$ are obtained for the $\gamma$ model --- a version of Eliashberg theory {in which  the effective electron-electron 
interaction is proportional to}
{$(g/|\omega_n-\omega_m|)^{\gamma}$}, {where $\omega_n-\omega_m$ is the transferred Matsubara frequency,}
{$g>0$ a reference energy, and $\gamma>0$ a parameter.}
 The rigorous lower bounds are based on a variational principle that {identifies $(2\pi T_c/g)^\gamma$ 
with the largest (positive) eigenvalue $\mathfrak{g}(\gamma)$ of an explicitly constructed compact, 
self-adjoint operator $\mathfrak{G}(\gamma)$.}
 These lower bounds form an increasing sequence that converges to $T_c(g,\gamma)$.
 The upper bound on $T_c(g,\gamma)$ is based on fixed point theory, proving 
linear stability of the normal state for $T$ larger than the upper bound on $T_c(g,\gamma)$.
 \vspace{0.3truecm}
}

\thispagestyle{empty}

\vfill\vfill
\hrule
\medskip
\noindent
\copyright(2025) 
\small{The authors. Reproduction of this preprint, in its entirety, is permitted for non-commercial purposes only.}
\newpage

\section{Introduction}\vspace{-10pt}

 Superconductivity is a property of certain materials to conduct electric currents without dissipation.
 While in the  metallic (normal) state   the resistance decreases gradually with decreasing 
temperature, the superconducting state sets in abruptly at a certain critical (transition) 
temperature $T_c$ where the electrical resistance of the material suddenly drops to zero.
 Another defining characteristic of the superconducting
state is the expulsion of magnetic flux from its interior, which, in particular, allows superconductors 
to levitate in applied magnetic fields. 

In this paper we initiate a rigorous inquiry into $T_c$, that will be continued in several follow-up papers, each 
one building on the results of the previous one, \blue{yet written so that they stand in their own right --- modulo the 
proofs of some of the results that will be referred to}.
 In this section we present what is meant as a ``master introduction'' to our project:
 After briefly recalling the empirical, and then the theoretical state of affairs, we offer our mathematical critique
and propose our ``plan of attack'' for conquering this list of mathematical problems, followed by a summary of the 
results obtained in the present paper. 
\vspace{-10pt}

\subsection{Empirical state of affairs}

Since the discovery of superconductivity in mercury in 1911 by {H. Kamerlingh Onnes} \cite{onnes}, there has been a sustained effort, 
particularly intense in recent years, to discover superconductors with higher values of $T_c$. 
One of the main goals is to achieve superconductivity at room temperatures, which holds the promise of revolutionary 
technological applications. 
The critical temperatures of elemental superconductors, such as mercury, aluminum, and niobium, do not exceed 10~K,
though, and conventional superconducting compounds reach values of $T_c$ higher than that
only by less than an order of magnitude, up to 39~K in MgB$_2$ \cite{AllenMitrovic,Carbotte,allen-handbook,kong2,TroyanETal}.

In 1986, high-temperature superconductors were discovered  by  G. Bednorz and K. A. M\"uller \cite{BM}. 
These are materials with critical temperatures above 77~K, the boiling point of liquid nitrogen at ambient pressure. 
A notable example is the cuprate of mercury, barium, and calcium with $T_c$ 
above 130~K at ambient pressure \cite{SCGO}.
  Moreover, much higher values of $T_c$ were discovered, albeit at very high pressures, in entirely different types
of materials -- metallic hydrogen and hydride compounds. 
The current verified record belongs to LaH$_{10}$, (La,Y)H$_{10}$\cite{drozdov5,Somayazulu,sun}, (La, Sc)H$_{10}$\cite{dima}, and other ternary compounds containing lanthanum with  $T_c= 250$~K.

\subsection{Theoretical state of affairs}

 No compelling quantum-statistical physics explanation of the phenomenon of superconductivity has yet been supplied
that would work with the genuine quantum-mechanical hamiltonian $H_{\mbox{\tiny{Z,N}}}$ of $N$ nuclei
and of $NZ$ spin-$\frac12$ electrons, with electrical Coulomb interactions between all the charged particles, and reasonable 
spin-spin interaction among the electrons; here, $Z$ would have to be the atomic number of an elemental 
superconductor such as Hg or Pb.
 To show that the canonical free energy 
$- k_{\mbox{\tiny{B}}} T \ln \mathrm{tr}\, e^{- H_{\scriptscriptstyle{Z,N}}/k_{\mbox{\tiny{B}}} T}$ of 
an ensemble of such systems at a given density, taken per nucleus or per volume and considered in the thermodynamic limit, 
will at some $T_c>0$ feature a continuous phase transition between a normal (metallic) phase that exists in a right neighborhood
of $T_c$, and a superconducting phase in a left neighborhood of $T_c$, remains a truly hard challenge for current and future 
generations of theoretical physicists. 

 For the time being, the default point of departure continues to be the electron-phonon hamiltonian 
\begin{align}\label{eq:FH}
H^{\mbox{\tiny{F}}}
 = {\textstyle\sum\limits_{p,\sigma}}\xi(p) c^\dagger_{p,\sigma} c_{p,\sigma}
& + {\textstyle\sum\limits_q} \omega_0(q)b^\dagger_qb_q \\ \notag
& +\tfrac{1}{\sqrt{N}}
{\textstyle\sum\limits_{p,q,\sigma}}
\tfrac{\alpha(q)}{\sqrt{2M\omega_0(q)}} c^\dagger_{p+q,\sigma} c_{p,\sigma} \left[b^\dagger_{-q}+b_q\right]
\end{align}
for an ideal gas of spin-$\frac12$ fermions of momentum $p$ and spin projection $\sigma$
in interaction with an ideal gas of spin-$0$ bosons of momentum $q$, introduced in 1954 by {H. Fr\"ohlich} \cite{FroehlichH}.
 The fermions represent conduction band electrons, and the bosons represent phonons, {i.e.,} the quantized vibrational 
excitations of the ionic lattice \textsl{presumed} to be formed by the $N$ nuclei of mass $M$ and their bound electrons when
embedded in the neutralizing ideal gas of conduction band electrons.
 The momentum-energy dispersion relations $\xi(p)$ and $\omega_0(q)$, and the electron-phonon coupling function $\alpha(q)$, 
are theoretical input.
 The Fr\"ohlich hamiltonian still offers what can be called a microscopic approach to  {metals} in condensed matter
 {physics}, yet 
one that is reduced to what are thought to be the \textsl{effective} microscopic degrees of freedom, 
compared to $H_{\mbox{\tiny{Z,N}}}$.
 For rigorous results on its ground state, see \cite{FreericksLieb}.
 \vspace{-5pt}

\subsubsection{BCS theory in a nutshell}
 The first successful semi-microscopic theory of superconductivity was proposed by 
J. Bardeen, L. N. Cooper, and J. R. Schrieffer (BCS) in 1957 \cite{BCSa,BCSb}.
 The theoretical discovery of phonon-mediated electron-electron pairing into what was soon to become known as
 Cooper pairs suggested to BCS to drastically simplify $H^{\mbox{\tiny{F}}}$ further into
\begin{equation}
H^{\mbox{\tiny{BCS}}}
 = {\textstyle\sum\limits_{p,\sigma}}\xi(p) c^\dagger_{p,\sigma} c_{p,\sigma}
-
\tfrac{\lambda}{\nu_0}
{\textstyle\sum\limits_{\genfrac{}{}{0pt}{4}{p,q}{\genfrac{}{}{0pt}{4}{|\xi(p)|<\Omega_{\mbox{\tiny{cut}}}}
{|\xi(q)|< \Omega_{\mbox{\tiny{cut}}}}}}}
 c^\dagger_{q,\uparrow} c^\dagger_{-q,\downarrow} c_{p,\uparrow} c_{-p,\downarrow},
\end{equation}
a hamiltonian that operates only on the electron degrees of freedom. Here $\nu_0$ is the \textsl{bulk} density of states at the Fermi level.
 In going from $H^{\mbox{\tiny{F}}}$ to $H^{\mbox{\tiny{BCS}}}$ the two terms involving 
phonon degrees of freedom have been eliminated from the Fr\"ohlich hamiltonian in favor of
an effective interaction term involving pairs of Cooper pairs; only an electron-phonon coupling constant,
$\lambda$, remains as reminder of the role that phonons play in mediating the formation of Cooper pairs and their 
interaction. 
 A {spectral gap} $\Delta(0)$ is opened up  at the Fermi level by the formation of Cooper pairs.
 Thanks to the energy gap the Cooper pairs can support a {loss-less electrical supercurrent}, as long as this
current is not too large.

 In $\mathrm{tr} \exp\!\big(\!- H^{\mbox{\tiny{BCS}}}/k_{\mbox{\tiny{B}}} T\big)$
the quartic femionic term is itself simplified by replacing it by a symmetrized product of one quadratic factor
with the expected value of the other one -- invoking a mean-field approximation.
 This yields the {order parameter} $\Delta(T)>0$ as solution of what is known as the BCS gap equation,
\begin{equation}\label{eq:BCSgapEQ}
1 = \lambda \int_0^{\hbar\Omega_{\mbox{\tiny{cut}}}}
\frac{\tanh \left(\sqrt{\varepsilon^2 + \Delta^2}/2k_{\mbox{\tiny{B}}} T\right)}{\sqrt{\varepsilon^2 + \Delta^2}}d\varepsilon
\quad \mbox{for}\quad \lambda \ll 1,
\end{equation}
which exists for $T<T_c$, where $T_c$ is the unique $T$ value solving (\ref{eq:BCSgapEQ}) for when $\Delta=0$;
note that for $T>T_c$ r.h.s.(\ref{eq:BCSgapEQ}) is $<1$, and the BCS order parameter ceases to exist. 
 For $T\downarrow 0$ the order parameter $\Delta(T)$ becomes the {spectral gap} $\Delta(T=0)$.
 Measuring $T_c$ allows one to infer the \textsl{spectral gap} $\Delta(0) = A T_c$, where $A$ does not 
depend on $\lambda$ or $\Omega_{\mbox{\tiny{cut}}}$ and can be computed from the theory. 

 While the BCS theory \blue{of superconductivity caused by phonon-induced Cooper pairs}
has been very successful, earning its authors a Nobel Prize in physics, it leaves many things desired.
 In particular, the BCS hamiltonian invokes an ad-hoc frequency cutoff $\Omega_{\mbox{\tiny{cut}}}$, somewhere
near the Fermi energy of the ideal electron gas (in units of $\hbar=1$); without such a cutoff,
i.e. when $\Omega_{\mbox{\tiny{cut}}}\to\infty$, 
then r.h.s.(\ref{eq:BCSgapEQ}) diverges to $\infty$ for any $T>0$, $\Delta>0$, and $\lambda>0$.

 As a result, BCS theory is neither able to accurately determine the superconducting $T_c$, nor the coupling constant $\lambda$,
nor many other properties.   
 Moreover, it is  quantitatively valid only for weak electron-phonon interactions 
(the small $\lambda$ regime).\footnote{\blue{This criticism of the original BCS theory does not apply to 
a variation on the theme of BCS theory that was developed about 25 years ago \cite{BCDB} to predict the onset of 
\emph{superfluidity} in trapped ultracold fermion gases.
  The interactions between the fermionic atoms in such systems are very different from the phonon-mediated 
electron-electron interactions that cause Cooper pairs in superconductors. 
 In recent years this model has received much attention by mathematical physicists, see e.g.
\cite{HSa}, \cite{HSb}, \cite{LT}, \cite{HLR}, but these studies do not alleviate the criticism of 
the BCS theory of phonon-mediated superconductivity.}}
 \vspace{-5pt}

\subsubsection{Eliashberg theory in a nutshell}
G. M. Eliashberg \cite{Eliashberg} remedied these issues in 1960 by building on an earlier work of A. B. Migdal \cite{migdal}
on electron-phonon interactions in metals that also took $H^{\mbox{\tiny{F}}}$ as point of departure.
 Eliashberg theory extends BCS theory to stronger electron-phonon interactions and allows one to 
directly incorporate the actual phonon spectrum; for reviews, see \cite{AllenMitrovic,Carbotte,Marsiglio}.
 Moreover, E.-G. Moon and A. Chubukov recently introduced a modification of the standard Eliashberg model 
\cite{MoonChubukov} (see also \cite{ChubukovETal}, \darkred{\cite{AC}, \cite{WAC}, \cite{WAWC}}), which they named the $\gamma$ model 
and which seeks to describe superconductivity in systems close to quantum phase transitions, where the effective electron-electron 
interactions are mediated by collective bosonic excitations (fluctuations of the order parameter) instead of phonons.

 Eliashberg theory, like BCS theory, is obtained from the quantum statistical mechanics formalism through a mean-field 
approximation to the free-energy density in the thermodynamic limit; \blue{in addition to the original papers and the cited reviews,
see also \cite{YuzAltPRB}, and also \cite{HonSal} for a renormalization group argument.
 Yet, Eliashberg theory} is much more refined than BCS theory.
 In Eliashberg theory, the BCS gap equation (\ref{eq:BCSgapEQ}) is replaced by an infinite set of 
coupled nonlinear equations for the Eliashberg gap function $n\mapsto \Delta_n(T)\geq 0$, known jointly as the 
Eliashberg gap equation, from which the order parameter $\Delta(T) \geq 0$, and the energy gap $\Delta(0)$, can be extracted.
 The Eliashberg gap equation reads \blue{(cf. eq.(38) in \cite{YuzAltPRB})}
\begin{equation}\label{eq:EgapEQ}
\forall n\in\ZZ:\ 
\omega_n \Delta_n = \pi T \sum_{m\in\ZZ} \lambda_{n,m} \frac{\omega_n\Delta_m - \omega_m\Delta_n}{\sqrt{\omega_m^2+|\Delta_m|^2}}.
\end{equation}
 Here, the $\omega_k: = (2k+1)\pi {T}$ for $k\in \ZZ$ are fermionic Matsubara frequencies, 
and $\lambda_{n,m}\equiv \lambda(n-m)\equiv V(\omega_n-\omega_m)$ is  
a dimensionless effective electron-electron interaction, of which we are particularly interested in the following kinds.

In units where Boltzmann's constant $k_{\mbox{\tiny{B}}}=1$ and the reduced Planck constant $\hbar =1$,
the effective electron-electron interaction mediated by generally dispersive phonons is given by
\begin{equation}\label{eq:VphononGENERAL}
V_\mathrm{ph}(\omega_n-\omega_m)
:= 2 \int_0^\infty \frac{\alpha^2\! F(\omega)\omega}{\omega^2+(\omega_n-\omega_m)^2} d\omega.
\end{equation}
 Here, \blue{$\omega\mapsto \alpha^2\!F(\omega)\geq 0$} is the electron-phonon spectral
 function,\footnote{We note that in this standard notation, $\alpha^2\!F$ is a compound symbol.
 This notation has its historical roots in early models with a function $\alpha^2(\omega)\geq 0$ and phonon
density of states $F(\omega)\geq 0$ that where defined separately. 
Subsequently it was realized that the combination $\alpha^2\!F$ generalizes to situations in which 
the individual functions are not separately defined.
 It would be prudent to just write $f(\omega)$ (say), yet here we follow the tradition of the superconductivity literature.}
defined as the thermodynamic limit of a sum of Dirac $\delta$ measures with positive coefficients that are concentrated 
on a discrete set of frequencies $\Omega_k\in (0,\overline{\Omega}]$ with $\overline{\Omega}<\infty$, and \blue{which converges
in the sense of distributions to a Lebesgue integrable function} $\alpha^2\!F(\omega)$ \blue{($\propto\omega^2$ for small $\omega$)}.
 The quantity $V_\mathrm{ph}(0)=:\lambda$ is the dimensionless electron-phonon coupling constant of the theory, 
\begin{equation}\label{eq:lambda}
\lambda = 2 \int_0^\infty \frac{\alpha^2\! F(\omega)}{\omega} d\omega.
\end{equation}
 Note that our $\lambda$ is the \textsl{standard} (renormalized) dimensionless electron-phonon coupling constant 
of the Eliashberg theory; cf.~\cite{AllenMitrovic,Carbotte,AllenDynes}.

 Since $\omega_n-\omega_m = (n-m)2\pi T $, it has also become customary to use the notation $\lambda(n-m)$ instead of
$V_\mathrm{ph}(\omega_n-\omega_m)$, i.e.
\begin{equation}\label{eq:lambdaGENERAL}
\lambda(n-m):= 2 \int_0^\infty \frac{\alpha^2\! F(\omega)\omega}{\omega^2+4\pi^2T^2(n-m)^2} d\omega.
\end{equation}
 With this convention, the coupling constant $\lambda\equiv \lambda(0)$.
 In order to avoid any ambiguous statements, 
we will use $\lambda$ exclusively to mean the coupling constant (\ref{eq:lambda}), and not as abbreviation for the map
$j\mapsto \lambda(j)$, with $j\in\ZZ$, as defined by (\ref{eq:lambdaGENERAL}) --- something that is sometimes 
done in the superconductivity literature, unfortunately.
 Whenever we invoke (\ref{eq:lambdaGENERAL}) we will use the notation $\lambda(j)$, with $j\in \ZZ$ (or $j\in\NN$).

 For materials predominantly featuring what's known as \emph{optical phonons} 
it suffices to work with dispersionless (Einstein) phonons of renormalized frequency $\Omega$, for which
\begin{equation}\label{eq:EinsteinAlphaF}
 \alpha^2\! F(\omega)\Big|^{}_{\mbox{\tiny{Ein}}} := \tfrac{g^2}{2\Omega}\delta (\omega-\Omega).
\end{equation}
{This model is {sometimes} called the Holstein model, after \cite{Holstein,Holstein1}, 
but one should keep in mind that in the original Holstein model {it is the bare phonons that} 
are dispersionless, while here the physical (renormalized) phonons are. 
With this understanding, we will interchangeably refer to this model as the Holstein or the Einstein phonon model.}
 The dimensionless effective electron-electron interaction mediated by Einstein phonons therefore reads
\begin{equation}\label{eq:Vphonon}
V_\mathrm{ph}(\omega_n-\omega_m)\Big|^{}_{\mbox{\tiny{Ein}}} =\frac{g^2}{\Omega^2+(\omega_n-\omega_m)^2}.
\end{equation}
 In this realization of Eliashberg theory
the dimensionless electron-phonon coupling constant is given in terms of $g$ and $\Omega$, as
\begin{equation}
\lambda=\frac{g^2}{\Omega^2}.
\end{equation}

 We remark that with $\lambda=\lambda(0)$ well defined, it appears to not enter the Eliashberg gap equation
(\ref{eq:EgapEQ}) because the $m=n$ term in the summation manifestly vanishes identically.
 Yet, appearances are misleading. 
 This is clear in the Holstein model, where $g$ can be eliminated in favor of $\lambda$, yielding
\begin{equation}\label{eq:VphononALT}
V_\mathrm{ph}(\omega_n-\omega_m)\Big|^{}_{\mbox{\tiny{Ein}}} =\lambda\frac{\Omega^2}{\Omega^2+(\omega_n-\omega_m)^2}.
\end{equation}
 Similarly, we can rewrite the effective electron-electron interaction mediated by generally dispersive phonons as 
\begin{equation}\label{eq:VphononGENERALalt}
V_\mathrm{ph}(\omega_n-\omega_m)
=: \lambda \int_0^\infty \frac{ \omega^2}{\omega^2+(\omega_n-\omega_m)^2} P(d\omega),
\end{equation}
where $P(d\omega)$ is a non-negative normalized measure, integrating to unity, having a 
Radon--Nikodym derivative $P^\prime(\omega)\propto\omega$ for small $\omega$.

 In the $\gamma$ model the dimensionless effective electron-electron interaction by definition is
 \begin{equation}\label{eq:Vgamma}
V_\gamma(\omega_n-\omega_m) :=\frac{g^\gamma}{|\omega_n-\omega_m|^\gamma},\quad n\neq m,
\end{equation}
and $V_\gamma(0):=0$; here, $\gamma$ can in principle take arbitrary positive values. 
 In particular, it has been argued that  $\gamma=\frac13$, respectively $\gamma = \frac12$, for two-dimensional nematic,
respectively magnetic quantum critical points, that $\gamma\approx \frac{7}{10}$ for a spin-liquid model for the cuprates, 
and $\gamma=1$ for pairing mediated by an undamped propagating boson in two dimensions; 
{see \cite{ChubukovETal} and references therein for details.}
  
 We remark that (\ref{eq:Vgamma}) in the special case $\gamma=2$ is obtained for $m\neq n$ by letting $\Omega\searrow 0$ in
(\ref{eq:Vphonon}). 
 Thus the $\gamma$ model might be thought of as a generalization of a 
$\lambda\to\infty$ limit of the rescaled Holstein model,
the rescaling being expressed through the finite coupling constant $g$ being used in place of the infinite $\lambda$. 
Incidentally, we emphasize that in this series of papers we study bounds on $T_c$ \textsl{within} 
the above models, as opposed to bounds stemming from the limits of their applicability to actual physical systems 
\cite{YuzAltPRB2,SemenokBorisEmil,EsterlisETal,EKS,CAEK,YuzKieAltPRB,Tra,Sad,YuzAltPat}. 
In particular, even though $\lam$ in real metals cannot exceed a critical value 
{$\lam_c\in [3,4]$}, the study of the $\lam\to\infty$ limit is still quite useful; e.g., it provides 
a simple asymptotic upper bound on $T_c$ that presumably is valid for all $\lam$. 
 Note also that the suitably rescaled limits $\lam\to 0$ (yielding the BCS theory) and $\lam\to\infty$  are the two universal 
limits of the Eliashberg theory \cite{AllenMitrovic,Carbotte,AllenDynes,Combescot,YuzAltPRB2}, in the
sense that the theory becomes independent of the phonon spectrum and other microscopic details.
 Some of its key features become most apparent, then.

 Coming back to the Eliashberg gap equation, a manifestly obvious solution is $\Delta_n\equiv 0\, \forall\, n\in\ZZ$;
it maps into a vanishing order parameter $\Delta(T)\equiv 0$. 
 Since the order parameter vanishes above $T_c$ when the {material} is in its normal state, the trivial solution
$\Delta_n\equiv 0\, \forall\, n\in\ZZ$ corresponds to the normal state, for all $T>0$.

 However, the normal state has been numerically found to be linearly stable, in the sense of locally minimizing the
free-energy density of Eliashberg theory, only for temperatures above a critical temperature $T_c$, and unstable below $T_c$.
 Furthermore, nontrivial solutions of the nonlinear Eliashberg gap equation that can be identified with a superconducting state
have been found numerically only for $T<T_c$.
 Most importantly, such types of numerical evaluations of the Eliashberg equations have accurately predicted $T_c$ for a 
broad range of superconductors, including the hydrides \cite{AllenMitrovic,Carbotte,AllenDynes,TroyanETal}. 
\vspace{-10pt}

\subsection{Mathematical critique and goals}\vspace{-5pt}

  While the advantages of the Eliashberg theory over the BCS theory for the treatment of electron-phonon superconductors
are not seriously in doubt,
it should be noted that this success story is based primarily on a practical mix of ad-hoc approximations
to the Eliashberg equations and heuristic assessments of when the computational schemes appear to have converged.
 For mathematically minded readers it is rather difficult to appreciate what is actually being accomplished in
standard references like \cite{AllenDynes,BergmannRainer} that claim to accurately compute $T_c$,
when in \cite{BergmannRainer} one also reads that ``The frequency summation in the [linearized]\footnote{Inserted 
by the present authors.} 
gap equation (2) should be done up to a frequency of the order of the band width. 
For technical reasons one only does the summation up to a cut-off frequency somewhat larger than the maximum phonon frequency.''
 This sounds very much like a BCS-type frequency cutoff $\Omega_{\mbox{\tiny{cut}}}$ having re-entered through a backdoor.
 Note that \cite{AllenDynes} start directly with Eq.~(7) of \cite{BergmannRainer}, also their Eq.~(7), a judiciously 
truncated version of the linearized Eliashberg gap equations.\footnote{\blue{In
Eq.~(7) of \cite{BergmannRainer} the kernel $\lambda(n-m)$ is shifted by the addition of an $N$-dependent 
``reduced Coulomb repulsion parameter'' $\mu^*(N)$.
 We emphasize that no such Coulomb term can be extracted from the Fr\"ohlich hamiltonian (\ref{eq:FH}), and so it has
no place in Eliashberg theory as derived from (\ref{eq:FH}).
 It is ``smuggelt'' into the truncated equations using some hand-waving physical reasoning,
basically for the purpose of obtaining better fits to experimental data.
 Note that no non-zero constant can be added to $\lambda(n-m)$ in Eliashberg theory because the summations over 
all Matsubara frequencies would diverge, then.}}
 Numerical evaluations of such truncated linearized equations in themselves cannot address the question whether a critical
temperature $T_c$ is actually well-defined in terms of the untruncated linearized Eliashberg gap equations, and even supposing 
so, whether this $T_c$ satisfies the thermodynamical definition.
 Even then, it is not a-priori clear whether $T_c$ can be accurately computed by the approximation through truncation.

 Put differently, in Eliashberg theory (here understood to include the standard Eliashberg model as well as the $\gamma$ model), 
$T_c$ is thermodynamically defined.
 If the theory predicts a thermodynamically well-defined critical temperature $T_c$, as function of $\lambda$
for the standard Eliashberg model, respectively as function of $\gamma$ for the  $\gamma$ model,
then $T_c$ has to be the borderline temperature between a connected high-temperature regime 
where the normal state is the thermal equilibrium state, and a connected low-temperature regime where a superconducting
state is the thermal equilibrium state.
 If this is the case, and if the phase transition at $T_c$ is a continuous phase transition, 
then the normal state is linearly stable against superconducting perturbations for temperatures above $T_c$, and 
unstable against such perturbations at temperatures below $T_c$.
  However, we are not aware of any rigorous arguments in the literature that this is what the theory predicts.  
The only rigorous result that we are aware of
states the \textsl{absence of odd-$f$ superconductivity} \darkred{for certain model realizations of} Eliashberg theory \cite{LangmannETal};
\darkred{odd-$f$ superconductivity has been reported to feature in the \emph{extended $\gamma$ model} \cite{WZAC}, though.}

 Instead, in the traditionally cited standard literature \cite{AllenDynes}, reviewed in \cite{AllenMitrovic}, 
it has been claimed that $T_c$ can be approximated from below arbitrarily precisely by determining when the largest eigenvalue 
of a real-symmetric $N\times N$ truncation of an operator associated with the linearized Eliashberg gap equation
vanishes,\footnote{Intuitively, think of the eigenvalues as rates  for exponentially time-dependent eigenmodes. 
 If they are all negative, the eigenmodes decay in the course of time and the normal state is linearly stable. 
 But if there is even only a single positive eigenvalue, then there is an exponentially growing mode, and the normal 
state is unstable.}
as long as $N$ is ``large enough.'' 
 Numerical studies involving up to $N=64$ Matsubara frequencies in \cite{AllenDynes}
 have been offered in support of this narrative, yet a mathematical vindication has been missing.

 More explicitly, it is easy to understand that the largest eigenvalue of a family of symmetric $N\times N$ truncations of 
(let's call it) the ``linearized Eliashberg gap operator'' grows monotonically with $N$, since these truncations are projections
onto an expanding family of $N$-dimensional subspaces of the Hilbert space on which the linearized Eliashberg gap operator acts.
 Yet, without a suitable monotonic dependence of this eigenvalue on $T$ it does not follow that increasing $N$ would lead
 to an increasing sequence of lower approximations to $T_c$.
 True, for a truncation of this eigenvalue problem to a single (the first positive) Matsubara frequency it is manifest
that the only eigenvalue of the pertinent $1\times 1$ matrix depends monotonically on $T$ and changes sign once when $T$ is varied 
(cf. the text surrounding equations (18)--(21) in \cite{AllenDynes}). 
 However, when the first $N>1$ positive Matsubara frequencies are involved in the
truncation the largest eigenvalue of the pertinent $N\times N$ matrix does not manifestly depend suitably monotonically on $T$, 
claims to the contrary in \cite{AllenDynes} notwithstanding.
 While the numerical studies in \cite{AllenDynes} apparently suggest this monotonicity also for $N\in\{2,4,8,16,32,64\}$, 
this cannot replace a proof that holds for all $N$, and for other values of the coupling constant $\lambda$ than those covered
in the numerical studies of \cite{AllenDynes}. 
 
 Also lacking in the literature has been an argument for why the eigenmode associated with 
a largest eigenvalue of the Eliashberg gap operator  of the normal state {corresponds to the superconducting}
{minimizer of the free energy functional in Eliashberg theory.}
 For this to be the case one needs to have  $\Delta_n >0\, \forall\, n\in\ZZ$~\cite{YuzAltPRB}, something that cannot be taken 
for granted just because the corresponding numerically computed mode in the $N\times N$ truncations with
particular choices of $N$ and $\lambda$ feature the analogous positivity.  

 One could easily go on and list further points of criticism, but the above list of problems already is too
long to {be addressed} rigorously in a single paper. 
 It paves the ground for explaining what we plan to do in a series of papers, of which the present one features part I.

 The present paper is devoted to a study of $T_c$ as defined by the linearized Eliashberg gap equations for the $\gamma$ model.
 The results obtained here are of interest in their own right, yet they {also} will be important ingredients for the 
follow-up part II paper, which deals with the linearized Eliashberg gap equations for the standard version of Eliashberg theory 
in which the effective electron-electron interactions are mediated by generally dispersive phonons.
 In our part III paper we in turn will specialize our results of paper II to their non-dispersive limit, {sometimes
called the} Holstein model, where much more detailed results can be obtained that also are of interest in their own right.
 Recall that the special case $\gamma=2$ of the $\gamma$ model captures the $\lambda\to\infty$ limit
{of a properly rescaled} Holstein model. 

 A study of the nonlinear Eliashberg gap equations is reserved for yet another publication.

\subsection{Synopsis of our results for $T_c(g,\gamma)$}

{The $\gamma$ model features two parameters, $g$ and $\gamma$.}
 We rigorously study the linear stability of the normal state, i.e. of the trivial solution
of the linearized Eliashberg gap equation, against small perturbations, {as a function of $g$ and $\gamma$.
 We find that for each $\gamma>0$ and $g>0$} there is a critical $T_c(g,\gamma)$ such that the normal state is linearly stable
when $T>T_c(g,\gamma)$ and unstable against superconducting perturbations when $T<T_c(g,\gamma)$.
 Thus $T_c(g,\gamma)$ is well-defined in the sense of linear stability analysis.

 We characterize $T_c(g,\gamma)$ by a variational problem for the largest eigenvalue
$\mathfrak{g}\bigl(\gamma\bigr)$ of a $\gamma$-dependent self-adjoint, compact operator $\mathfrak{G}(\gamma)$ 
on the Hilbert space of square summable sequences, introduced explicitly in section~4.
 The critical temperature $T_c(g,\gamma)$ is given by
\begin{align}
\label{eq:TcGAMMA}
 T_c(g,\gamma)
 =\, \tfrac{g}{2\pi} \bigl[\mathfrak{g}\bigl(\gamma\bigr) \bigr]^{\frac1\gamma}.
\end{align}

{
Formula (\ref{eq:TcGAMMA}) for $ T_c(g,\gamma)$ reveals that $g$ plays a rather simple role in the analysis. 
In particular, for many purposes we may simply think of $T_c$ as measured in units of $g$, 
which is equivalent to setting $g = 1$, and  ease the notation by simply writing $T_c(\gamma)$ for $T_c(1,\gamma)$.
 For other purposes, such as comparisons of the $\gamma$ model results with results of other models that do not
feature $g$, it is of advantage to keep $g$ as an overall factor, and so we will continue to do so for now.
 We will explicitly alert the reader when we switch to setting $g=1$.}

 We also show that the {eigenmode for the largest eigenvalue} of $\mathfrak{G}(\gamma)$ is the eigenmode for the 
spectral radius of a (non-symmetric) operator that maps the positive cone of $\ell^2$ into itself. 
 With the input of the Krein--Rutman theorem {that eigenmode} can be identified as  {corresponding to the 
superconducting} {minimizer of the condensation energy functional.}
 
 The compactness of  $\mathfrak{G}(\gamma)$ on the Hilbert space $\ell^2$ implies that  $\mathfrak{G}(\gamma)$ 
can be arbitrarily accurately approximated through \blue{truncations $\mathfrak{G}^{(N)}(\gamma)$ to $N$}-dimensional 
subspaces, by choosing their dimensions large enough. 
\blue{Each $\mathfrak{G}^{(N)}(\gamma)$ is a real-symmetric $N\times N$ matrix that is the upper left $N\times N$ block
of the ensuing matrix $\mathfrak{G}^{(N+1)}(\gamma)$.
 We give $\mathfrak{G}^{(4)}(\gamma)$ explicitly below, featuring 
$\mathfrak{G}^{(N)}(\gamma)$ for $N\in\{1,2,3\}$ as submatrices, and revealing the pattern to construct 
$\mathfrak{G}^{(N)}(\gamma)$ for any $N\in\NN$.}
 Our truncations lead to a strictly increasing infinite sequence of lower bounds on $T_c(g,\gamma)$ that
converges to $T_c(g,\gamma)$, viz.
\begin{align}
\label{eq:TcLOWERbounds}
\hspace{-1truecm}
 T_c^{(1)}(g,\gamma) <  \cdots <  T_c^{(n)}(g,\gamma) < T_c^{(n+1)}(g,\gamma) <\cdots & < T_c^{(\infty)}(g,\gamma)\\
\hspace{-1truecm}                   & = T_c(g,\gamma).
\end{align}
 The first four of these bounds can be computed explicitly. 
 They read
\begin{align}
\label{eq:TcLOWERboundsONE}
\hspace{-1truecm}
 T_c^{(1)}(g,\gamma) & = \tfrac{g}{2\pi}; \\
\label{eq:TcLOWERboundsTWO}
\hspace{-1truecm}
 T_c^{(2)}(g,\gamma) & =  \tfrac{g}{2\pi}\!
\left[\tfrac16\!\left(\!\! 1 \!+\!\tfrac{1}{3^\gamma} \!+\! 
\sqrt{\!\big(1 + \tfrac{1}{3^\gamma}\big)^2\! + \!
12\left(\!\! \big(1 + \tfrac{1}{2^\gamma}\big)^2\!\! +\! 
2 - \tfrac{1}{3^\gamma} \right)\!}\right)\!\right]^{\!\frac1\gamma}\!\!\!\!;\!\! \\
\label{eq:TcLOWERboundsTHREE}
\hspace{-1truecm}
T_c^{(3)}(g,\gamma) & = \tfrac{g}{2\pi}
\left(\!  \textstyle{\frac{r(\gamma)}{3}  +
2 \sqrt{\frac{p(\gamma)}{3}}\cos \left[\frac13\arccos\left(\!\frac{q(\gamma)}{2}\sqrt{\!\Big(\frac{3}{p(\gamma)}\Big)^{\!{}_3}}\right)\!
\right] } \right)^\frac1\gamma
\end{align}
with $p,q,r$ given by 
\begin{align}
\label{eq:gamma3x3Kp}
\hspace{-.8truecm}
p(\gamma) = & \tfrac13\big({\rm tr}\,\mathfrak{G}^{(3)}(\gamma)\big)^2- {\rm tr\, adj}\, \mathfrak{G}^{(3)}(\gamma),\\
\label{eq:gamma3x3Kq}
\hspace{-.8truecm}
q(\gamma) = & \tfrac{2}{27}\big({\rm tr}\,\mathfrak{G}^{(3)}(\gamma)\big)^3 
- \tfrac13 {\rm tr}\,\mathfrak{G}^{(3)}(\gamma)\;{\rm tr\, adj}\,\mathfrak{G}^{(3)}(\gamma)
 + \det\mathfrak{G}^{(3)}(\gamma),\!\! \\
\label{eq:gamma3x3Kr}
\hspace{-.8truecm}
r(\gamma) = & {\rm tr}\,\mathfrak{G}^{(3)}(\gamma) ,
 \end{align}
where $\mathfrak{G}^{(3)}$ is the upper left $3\times 3$ block of the matrix $\mathfrak{G}^{(4)}$ given below.
 The trace, the trace of the adjugate, and the determinant of $\mathfrak{G}^{(3)}$ are given in Appendix~B.2.
 At last, and temporarily suppressing the dependence on $\gamma$ of the coefficients of the characteristic polynomial 
$\det \big(\eta\mathfrak{I} - \mathfrak{G}^{(4)}(\gamma)\big)=\eta^4 +A(\gamma)\eta^3+B(\gamma)\eta^2+C(\gamma)\eta+D(\gamma)$,
we have
\begin{align}\label{eq:TcLOWERboundsFOUR}
\hspace{-1truecm}
T_c^{(4)}(g,\gamma) = 
\tfrac{g}{2\pi}\!\left[\!-\tfrac14 A +\! \sqrt{\tfrac12 Z}
+\!\sqrt{\!\tfrac{3}{16} A^2 -\tfrac12 B - \tfrac12 Z +\tfrac{A^3 -4AB + 8C}{16\sqrt{2Z}}}
\right]^{\frac1\gamma}\!,
\end{align}
where  $Z(\gamma)$ is a positive zero of the so-called \textsl{resolvent cubic} associated with the characteristic polynomial 
$\det \big(\eta\mathfrak{I} - \mathfrak{G}^{(4)}(\gamma)\big)$,
given by (temporarily suspending displaying the dependence on $\gamma$ again)
\begin{equation}
\label{eq:Zdef}
Z = \tfrac13 \Big[ \sqrt{Y} \cos\Big(\tfrac13 \arccos \tfrac{X}{2\sqrt{Y^3}}\Big) - B + \tfrac38 A^2\Big],
\end{equation}
with  
\begin{align}
\label{eq:Xdef}
X =&  2B^3 - 9ABC+27C^2 + 27A^2D -72BD,\\
\label{eq:Ydef}
Y =& B^2-3AC+12D,
\end{align}
where
\begin{align}
\label{eq:Adef}
A &= - {\rm tr}\, \mathfrak{G}^{(4)},\\
\label{eq:Bdef}
B &= \tfrac12\Big( \big({\rm tr}\, \mathfrak{G}^{(4)}\big)^2 - {\rm tr}\, \big({\mathfrak{G}^{(4)}}\big)^2\Big),\\
\label{eq:Cdef}
C &= -\tfrac16 \Big( 
\big({\rm tr}\, \mathfrak{G}^{(4)}\big)^3 - 3 {\rm tr}\, \big({\mathfrak{G}^{(4)}}\big)^2 \big({\rm tr}\, \mathfrak{G}^{(4)}\big)
+ 2{\rm tr}\, \big({\mathfrak{G}^{(4)}}\big)^3\Big),\\
\label{eq:Ddef}
D &= \det {\mathfrak{G}^{(4)}},
\end{align}
and with (restoring the dependence on $\gamma$)
\begin{align}\label{eq:Hfour}
&\hspace{-1truecm}
\mathfrak{G}^{(4)}(\gamma) = \\ \notag
&\hspace{-1.5truecm}
 {\begin{pmatrix}
   1
&  \frac{1}{\sqrt{3}}\left[\frac{1}{2^\gamma }  + 1 \right]
&  \frac{1}{\sqrt{5}}\left[\frac{1}{3^\gamma }  + \frac{1}{2^\gamma }\right]  
&  \frac{1}{\sqrt{7}}\left[\frac{1}{4^\gamma }  + \frac{1}{3^\gamma }\right]   \\
 \frac{1}{\sqrt{3}}\left[\frac{1}{2^\gamma }  + 1 \right]
&  \frac{1}{3}\left[\frac{1}{3^\gamma }         - 2  \right]
&  \frac{1}{\sqrt{15}}\left[\frac{1}{4^\gamma } + 1 \right] 
&  \frac{1}{\sqrt{21}}\left[\frac{1}{5^\gamma } + \frac{1}{2^\gamma } \right]  \\ 
 \frac{1}{\sqrt{5}}\left[\frac{1}{3^\gamma }  + \frac{1}{2^\gamma }\right] 
&  \frac{1}{\sqrt{15}}\left[\frac{1}{4^\gamma } + 1 \right] 
&  \frac{1}{5}\left[\frac{1}{5^\gamma }  -  \frac{2}{2^\gamma } - 2 \right] 
&  \frac{1}{\sqrt{35}}\left[\frac{1}{6^\gamma } + 1 \right]  \\
 \frac{1}{\sqrt{7}}\left[\frac{1}{4^\gamma }  + \frac{1}{3^\gamma }\right]  & 
 \frac{1}{\sqrt{21}}\left[\frac{1}{5^\gamma } + \frac{1}{2^\gamma } \right] & 
  \frac{1}{\sqrt{35}}\left[\frac{1}{6^\gamma } + 1 \right]  & 
 \frac{1}{7}\left[\frac{1}{7^\gamma }  - \frac{2}{3^\gamma } -  \frac{2}{2^\gamma } -  2 \right] 
 \\
\end{pmatrix}}\!.
\end{align}
\smallskip
 The trace of the first three powers of $\mathfrak{G}^{(4)}$, and its determinant, 
are given in Appendix~B.3.

 For all $N>4$ the lower bounds $T_c^{(N)}(g,\gamma)$ on $T_c(g,\gamma)$ have to be computed by numerical
approximation of (\ref{eq:TcGAMMA}).
 Practically, this can be done accurately if $N$ is not too large, and $\gamma$ not too small. 

 We also constructed an upper bound ${T}_c(g,\gamma) \leq {T}_c^*(g,\gamma)$, with 
\begin{equation}\label{eq:TcUPPERboundOmNULL}
{T}_c^*(g,\gamma) = 
\frac{g}{2\pi}\!\left[1+ 2\big((2^{1+\epsilon(\gamma)}\!-1) \zeta(1\!+\!\epsilon(\gamma)) \zeta(1+2\gamma-\epsilon(\gamma)) \big)^{\frac12}\right]^{\frac1\gamma}\!\!\!,\!\!\!
\end{equation}
where $\epsilon(\gamma) = \min\{\gamma,0.65\}$.
 In (\ref{eq:TcUPPERboundOmNULL}), $\zeta(s) := \sum_{n\in\NN} \frac{1}{n^{s}}$ for $s>1$ is the Riemann $\zeta$ function.
 When $\gamma\to \infty$, then  $T_c^*$ converges to the exact value $T_c(g,\infty) = \frac{g}{2\pi}$, while
it diverges $\sim \frac{g}{2\pi} \left({2}/{\gamma}\right)^\frac1\gamma$ when $\gamma\searrow 0$,
in agreement with the estimated behavior $T_c\sim C^\prime \left({C}/{\gamma}\right)^\frac1\gamma$ when $\gamma\searrow 0$ for 
certain $C,C^\prime>0$; see   \cite{WAAYC,YuzKieAltPRB}.

 In the formula for $T_c(g,\gamma)$, and in all our bounds on $T_c(g,\gamma)$, we have displayed $g$
explicitly for later convenience and reference.
 In much of what follows we will set $g=1$ and simply write $T_c(\gamma)$ for $T_c(1,\gamma)$, to ease the notation.

 \magenta{The next figure shows our lower bounds $T_c^{(N)}(\gamma)$ for $N\in\{1,...,4\}$, 
our upper bound $T_c^*(\gamma)$, and also a numerically computed approximation to 10 decimal places
precision of the lower bound $T_c^{(400)}(\gamma)$ for a sampling of $\gamma$ values between $0.4$ and $3.1$ spaced by steps 
of $0.1$; we also added half a dozen intermediate $\gamma$ values below 1 to fill in the larger gaps between the $T_c$ values 
that open up for smaller $\gamma$ values when these are equally spaced.
 We produced the $T_c^{(400)}(\gamma)$ data points using the Eigenvalues command of the LinearAlgebra package of Maple2021 to
compute the largest eigenvalue of $\mathfrak{G}^{(400)}$, then used the analog of (\ref{eq:TcGAMMA}) for this truncation of
the eigenvalue problem.
 We reproduced our numerical results with the alternate method described in \cite{YuzKieAltPRB}, finding agreement to 8 decimal places.
 We also checked the accuracy of the numerical Maple routine against the value of our exact lower bound $T_c^{(4)}(\gamma)$, finding
agreement to 9 decimal places. 
 This gives us confidence in the accuracy of the data points for $T_c^{(400)}(\gamma)$. }

\begin{figure}[H]
 \centering
\includegraphics[width=0.7\textwidth,scale=0.7]{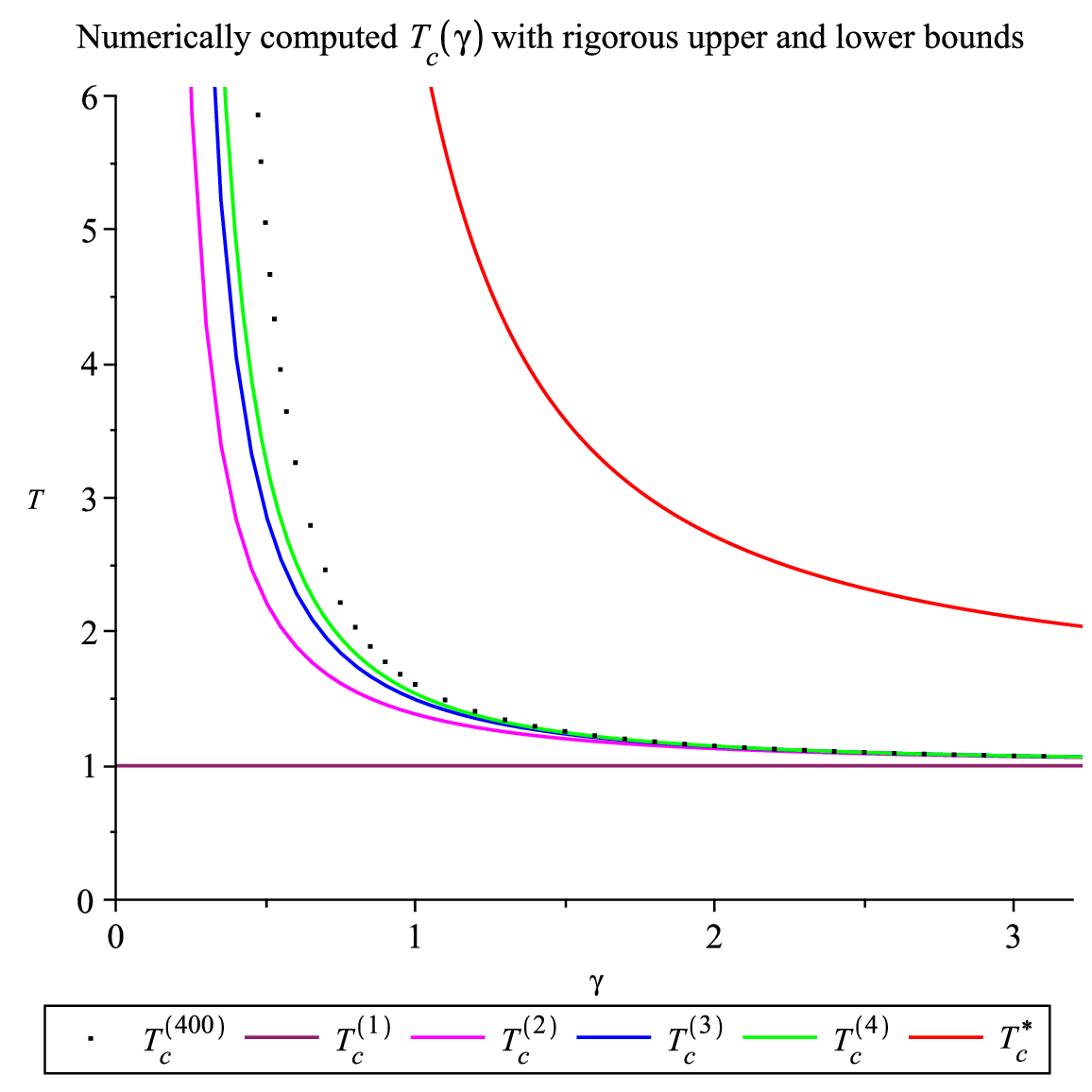} 
\vspace{-10pt}
\caption{\label{fig:TcBOUNDSgamma}
\footnotesize
Shown for the $\gamma$ model are data points for $T_c(\gamma)$, numerically computed with sequences truncated to 
the first 400 positive Matsubara frequencies.
Shown also are the graphs of the explicit lower bounds $ T_c^{(N)}(\gamma)$ with $N\in\{1,2,3,4\}$,
and the graph of the explicit upper bound $ T_c^*(\gamma)$. 
The temperature $T$ is in units of $g/2\pi$.}
\end{figure}

 Fig.~\ref{fig:TcBOUNDSgamma} reveals that \magenta{for $\gamma > 2$ our numerically computed data points 
with $N=400$ virtually lie on the $N=4$ graph, and for $1<\gamma<2$ they lie barely above that graph.
 This indicates a} rapid convergence \magenta{of the sequence of lower bounds} to the limiting graph of $ T_c(\gamma)$ when $\gamma>1$.
 Due to the blow-up of $T_c(\gamma)$ when $\gamma\searrow 0$, accurate approximation requires larger $N$ for smaller $\gamma$, visually
indicated by the fact that our numerically computed $T_c^{(400)}(\gamma)$ data lie higher and higher above $T_c^{(4)}(\gamma)$ the smaller 
$\gamma$ becomes.
 The plot of the explicitly computed upper bound $ T_c^*(\gamma)$ indicates how far at most the lower bounds could
be off of $T_c(\gamma)$. 
 As $\gamma\to\infty$, the upper and lower bounds converge to $\frac{1}{2\pi}$; \magenta{hence, so does $T_c(\gamma)$}. 

\begin{figure}[H]
 \centering
\includegraphics[width=0.7\textwidth,scale=0.7]{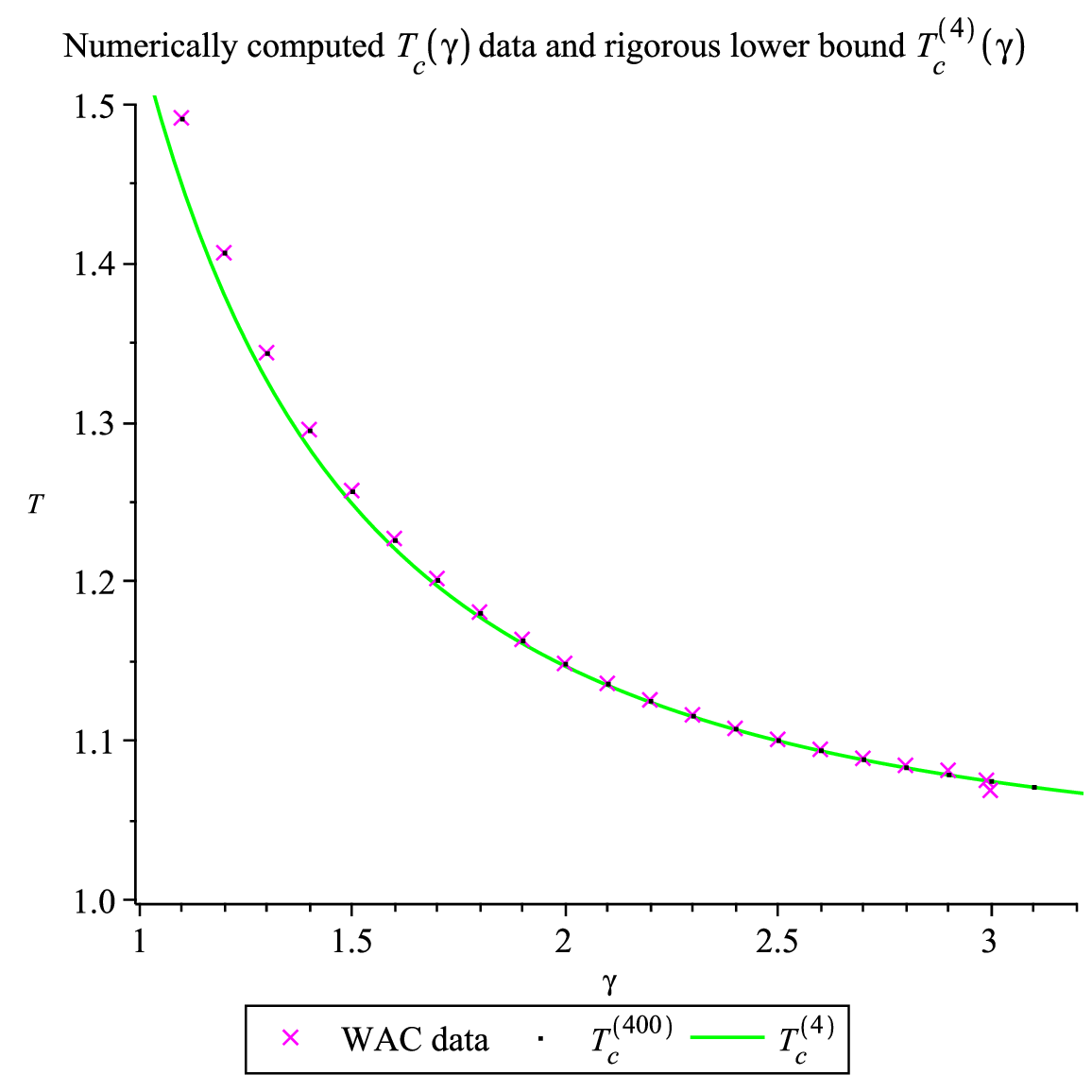}
\vspace{-10pt}
\caption{\label{fig:TcBOUNDSgammaZOOMin}
\footnotesize
A finer comparison of the rigorous lower bound $T_c^{(4)}(\gamma)$ and data points 
for $T_c(\gamma)$ computed with different algorithms; see main text.
Again, $T$ is in units of $g/2\pi$.}
\end{figure}

\darkred{It is also of interest to compare our explicit bounds with (approximate) $T_c(\gamma)$ data that have been computed 
numerically by other authors. 
 In Fig.~\ref{fig:TcBOUNDSgammaZOOMin} we show for the interval $1<\gamma<3.2$ the evaluation
of our explicit bound $T_c^{(4)}(\gamma)$, our numerically computed data for $T_c^{(400)}(\gamma)$,
and this time also the data for $T_c(\gamma)$ used to produce Fig.~1 in \cite{WAC} (here multiplied with $2\pi$
to match the units for $T_c$ in our Fig.~\ref{fig:TcBOUNDSgammaZOOMin}). 
 The truncation used in \cite{WAC} corresponds to $N=5000$ in our approach.
 Our upper bound $T_c^*(\gamma)$ is off the charts for Fig.~\ref{fig:TcBOUNDSgammaZOOMin}.}

\darkred{Fig.~\ref{fig:TcBOUNDSgammaZOOMin} reveals some interesting details. 
 For $\gamma$ values from $1.1$ to $2.9$, separated by steps of $0.1$, 
our numerical data points (dots) that have been computed with 400 positive Matsubaras
are barely distinguishable from the numerical data points of \cite{WAC} (crosses) 
that have been computed with the much larger number of Matsubaras (corresponding to 5000 positive ones).
 For $\gamma\geq 2$ those numerically computed $T_c(\gamma)$ data are actually barely distinguishable from the $N=4$ lower
approximation to $T_c(\gamma)$, indicating that the $N=4$ approximation is accurate enough for all $\gamma\geq 2$.
 We also computed $T_c^{(400)}(\gamma)$ for $\gamma=3$ and $\gamma=3.1$, whereas in \cite{WAC} $T_c(\gamma)$ data
have been computed for $\gamma=2.99$ and $\gamma=2.999$.
 Our numerical $N=400$ data for $\gamma\in \{3,3.1\}$ essentially agree with the $N=4$ lower bound,
but the same is true only for the $\gamma=2.99$ data point of \cite{WAC}, while the data point for $\gamma=2.999$ lies
below the $N=4$ lower bound on $T_c(\gamma)$. 
 This is indicative of some kind of singularity at $\gamma=3$ in the algorithm of \cite{WAC} that slows down its speed of convergence 
when $\gamma = 3$ is approached from below, so much so that for $\gamma=2.999$ the computed result has not the same
accuracy as the one for $\gamma=2.99$.}
 
\darkred{For $\gamma<2$ a narrow convex wedge opens up between the $N=4$ lower bound and the numerical data. 
 Our numerical data and those of \cite{WAC} continue to virtually agree also down to $\gamma=0.7$, but
for smaller $\gamma$ values numerical differences begin to show in the data.
 Yet, with the exception of the $\gamma = 2.999$ data 
point of \cite{WAC}, all numerical data satisfy our rigorous lower bound $T_c^{(4)}(\gamma)$.}

 The case $\gamma=2$ is of special interest, for it coincides with the $\Omega\to 0$ limit of the Eliashberg model 
with dispersionless (Einstein) phonons as given in (\ref{eq:Vphonon}) with {fixed $g$.}
 We thus state the explicit values of $T_c^{(N)}(2)$ for $N\in\{1,2,3,4\}$ {in units where $g=1$.}
 Our algebraic formulas yield
$T_c^{(1)}(2)=\frac{1}{2\pi}\approx 0.159154943$ and 
$T_c^{(2)}(2)=\frac{1}{2\pi} \frac{\sqrt{60 + 3\sqrt{13819}}}{18}\approx 0.1796160944$; 
we also have a closed transcendental expression for $T_c^{(3)}(2)$:
$$
\hspace{-1truecm}
T_c^{(3)}(2)
=\tfrac{1}{2\pi}\tfrac19 
\sqrt{\tfrac{1}{10^3}\left[-3284
 + \sqrt{a}\cos\left(\tfrac13\arccos\left(\tfrac{b}{\sqrt{a^3}}\right)\right)\right]},
$$
with $a \approx 12587996749$ and $b\approx 1101048630645068$;
hence, $T_c^{(3)}(2) \approx 0.1820383$.
 Also for $T_c^{(4)}(2)$ we have a closed transcendental expression, but this is already too unwieldy to be displayed
on two lines.
 Decimal expansion yields $T_c^{(4)}(2)\approx 0.1825137102$. 
Last not least, $T_c^*(2) \approx 0.3708637$, a factor $\approx 2.032$ larger than $T_c^{(4)}(2)$.

 Thus, truncation to merely $N=3$ Matsubara frequencies yields a lower bound on $T_c$ 
that agrees to three significant digits with the value $0.182 g$ reported in \cite{AllenDynes} (cf. their eq.(25)),
computed with a truncation of the Eliashberg gap equation to $N=64$ Matsubara frequencies.\footnote{Note that Allen 
and Dynes state $N+1=64$, because their $N$ count is shifted down by one unit compared to ours.}
 Yet, already the $N=4$ truncation yields the numerically better lower bound $T_c^{(4)}(2) = 0.1825...$, which 
rounds up to $T_c \approx 0.183$; cf. eq.(2.29) in \cite{Carbotte} where, however, \cite{AllenDynes} is given as reference.
 The $N=64$ computation of \cite{AllenDynes} should have produced this slightly larger rounded value than
the one they reported, but apparently it did not.

 Nowadays it is easy to reliably compute many more significant digits; cf. \cite{WAC}, \cite{WZAC}, \cite{YuzKieAltPRB}.
 We found that for $N$ beyond 200 the first 10 decimal places of the sequence of lower bounds 
$T_c^{(N)}(\gamma)$ have stabilized, yielding $T_c(2) = 0.1827262477...$; none of the displayed digits has been rounded.
\noindent

 The remainder of this paper is devoted to vindicating our rigorous results that we have stated in this subsection.
 We will work with the recent reformulation of Eliashberg theory in terms of a classical Bloch spin chain
 \cite{YuzAltPRB,YuzAltPRB2,YuzKieAltPRB}.


\section{The classical spin chain model}\label{sec:spinsCHAIN}
\vspace{-5pt}

 In \cite{YuzAltPRB} it was shown that the thermal equilibrium state in Migdal--Eliashberg theory can be represented
as a classical spin chain ${\bf S}\in ({\SP}^1)^{\ZZ}$.
 We let ${\bf S}_n \in {\SP}^1\subset {\RR}^2$ with $n\in{\ZZ}$ 
denote the $n$-th spin in the spin chain ${\bf S}$, written as a vector in ${\RR}^2$ of unit length.
 A spin chain ${\bf N}$ is associated with the {\textsl{normal state}} of the Migdal--Eliashberg theory, 
having $n$-th spin given by ${\bf N}_n := -{\bf N}_0\in  {\SP}^1\subset {\RR}^2$ for $n < 0$ and 
${\bf N}_n:={\bf N}_0$ for $n\geq 0$.
 Any other \textsl{admissible} spin chain satisfies the asymptotic conditions that, sufficiently fast,
${\bf S}_n \to {\bf N}_n$ when $n\to \infty$ and when $n\to - \infty$, where ``sufficiently fast'' is 
explained below.
 Moreover, admissible spin chains satisfy the symmetry relationship \textcolor{black}{that for all $n\in{\ZZ}$, 
 ${\bf N}_0\cdot {\bf S}_{-n} = - {\bf N}_0\cdot{\bf S}_{n-1}$ and 
 ${\bf K}_0\cdot {\bf S}_{-n} =  {\bf K}_0\cdot{\bf S}_{n-1}$, 
where ${\bf K}_0\in {\SP}^1\subset {\RR}^2$ is an arbitrary vector perpendicular to ${\bf N}_0$.}\footnote{\textcolor{black}{
In \cite{YuzKieAltPRB} we chose ${\bf N}_0$ to point along the $z$ axis and ${\bf K}_0$ along the $x$ axis.}}
 \textcolor{black}{This} reduces the problem to effective spin chains ${\bf S}\in  ({\SP}^1)^{{\NN}_0}$, with ${\NN}_0:={\NN}\cup\{0\}$.

 Admissibility is only a necessary condition for a spin chain to qualify as thermal equilibrium state. 
 For such a spin chain to actually represent a thermal equilibrium state it needs to minimize 
a certain thermodynamic functional. 
 In this paper we work with a normalized version of the functional given in \cite{YuzAltPRB}, 
known as the \textsl{condensation energy} of Eliashberg theory
 {[the difference between the grand (Landau) potentials of the superconducting and normal states]}.  
 Its spin chain representation reads
\begin{align}\label{eq:H}
 {H}({\bf S}|{\bf N}) : = &2\pi \textstyle\sum\limits_{n} \omega_n {\bf N}_0\cdot\big({\bf N}_n -{\bf S}_n \big)
  \\
\notag & + \pi^2 {T}
           \textstyle \sum\!\sum\limits_{\hskip-0.4truecm n\neq m}
\lambda_{n,m}^{} \left({\bf N}_n\cdot {\bf N}_m -{\bf S}_n\cdot{\bf S}_m \right),
\end{align}
where the summations still run over  ${\ZZ}$;
 we will switch to summations over ${\NN}_0$ in the next section.
Furthermore, $\lambda_{n,m}^{}$ could be any of the (dimensionless) positive spin-pair interaction kernels listed in the introduction.
 In this paper we work with 
\begin{equation}\label{eq:J}
\lambda_{n,m}^{} 
:= V_{\gamma}(\omega_n-\omega_m),
            \end{equation}
given in (\ref{eq:Vgamma}), {though in the following in units of $g=1$, for simplicity; 
recall our remark after (\ref{eq:TcGAMMA}).
 In our final formulas we restore $g$.}

 At last we are able to define \textsl{admissibility} of a spin chain ${\bf S}$ to mean, that 
after using their stipulated symmetry relations to convert all summations over negative Matsubara 
frequencies into summations over positive ones in the sum and the double sum in (\ref{eq:H}),
the resulting series converge absolutely.

 Based on many theoretical and numerical studies of its traditional formulation,
a general ``thermodynamic narrative''  for the Eliashberg theory has emerged 
\cite{AllenDynes,AllenMitrovic,Carbotte,Eliashberg,migdal,Marsiglio}.
It translates into the following mathematical conjecture for our classical spin chain $\gamma$ model:

\noindent
{\bf Conjecture 1}: \textsl{There is a critical temperature $T_c(\gamma) >0$
such that for temperatures  $T\geq T_c$, the spin chain of the normal {state} is the unique
minimizer of $H({\mathbf S}|{\mathbf N})$, whereas when $T<T_c$ a spin chain ${\bf S}\neq {\bf N}$ 
for a {\textsl{superconducting phase}} minimizes $H({\mathbf S}|{\mathbf N})$ uniquely 
up to an irrelevant gauge transformation (fixing of an overall phase).
 Moreover, the phase transition at $T_c$ from normal to superconductivity is continuous.} 
 \smallskip

 In this paper we take some steps toward the rigorous vindication of this conjecture, without 
aiming to give a proof of all its assertions.
 Instead, we assume that for each $\gamma>0$ and $T > 0$
there exists a minimizing spin chain of (\ref{eq:H}) in the set of admissible spin chains ${\mathbf S}$,
leaving a rigorous inquiry into the existence and uniqueness of minimizers to a future work.
 It is easy to see that any minimizer is a stationary point of $H({\mathbf S}|{\mathbf N})$ on the set of
admissible $\mathbf{S}$. 
 The present paper thus is exclusively concerned with the Euler--Lagrange equation for the stationary
points of $H({\mathbf S}|{\mathbf N})$. 
 Our discussion of the Euler--Lagrange equations, linearized about the normal {state} (trivial) solution, leads to
rigorous lower and upper bounds on $T_c$ --- defined as usual as the borderline value between the temperature regimes
of linear stability vs. instability of the spin chain of the normal state against ``superconducting perturbations.'' 
\vspace{-10pt}

\section{The Euler--Lagrange equations}\label{sec:EL}\vspace{-5pt}

 Recall that the symmetry relationship \textcolor{black}{that for all $n\in{\ZZ}$, 
 ${\bf N}_0\cdot {\bf S}_{-n} = - {\bf N}_0\cdot{\bf S}_{n-1}$ and 
 ${\bf K}_0\cdot {\bf S}_{-n} =  {\bf K}_0\cdot{\bf S}_{n-1}$, 
where ${\bf K}_0\in {\SP}^1\subset {\RR}^2$ is an arbitrary vector perpendicular to ${\bf N}_0$,}
allows us to work with effective spin chains ${\bf S}\in  ({\SP}^1)^{{\NN}_0}$, with ${\NN}_0:={\NN}\cup\{0\}$.
 Thus the Euler--Lagrange equations can be expressed in terms of summations over 
${\NN}_0:={\NN}\cup\{0\}$ instead of ${\ZZ}$.

 The restriction that the vectors ${\bf S}_n$ have unit length can easily be taken into account with the help
of infinitely many Lagrange multipliers. 
 Since $H({\bf S}|{\bf N})$ consists of linear and bilinear terms in ${\bf S}$, and since each unit-length
constraint ${\bf S}_n\cdot{\bf S}_n =1$ is quadratic, one obtains 
a \textsl{linear} Euler--Lagrange equations for infinitely many vectors ${\bf S}_n\in{\RR}^2$, 
$n\in{\NN}_0$, though constrained by infinitely many nonlinear conditions.

 Alternatively, the restriction that the vectors ${\bf S}_n$ have unit length can be implemented directly
by introducing an angle $\theta_n\in {\RR}/(2\pi{\ZZ})$ ($= [0,2\pi]$ with $2\pi$ and $0$ identified)
defined through ${\bf N}_0\cdot{\bf S}_n =: \cos \theta_n$ for all\footnote{If one also
introduces angles for spins with negative suffix by defining ${\bf N}_0\cdot{\bf S}_n =: \cos \theta_n$ for 
all $n\in -{\NN}$, a sequence of angles with non-negative 
suffix yields the angles with negative suffix as $\theta_{-1}=\pi-\theta_0$, $\theta_{-2}=\pi-\theta_1$, etc.,
thanks to the symmetry of ${\bf S}\in ({\SP}^1)^{{\ZZ}}$ with respect to the sign switch of the Matsubara frequencies.}
 $n\in {\NN}_0$.
 Setting ${H}({\bf S}|{\bf N}) =: {4\pi^2{T}}K_\gamma(\Theta)$ yields
\vspace{-.25truecm}
\begin{align}\label{eq:K}
\hspace{-.5truecm}
K_\gamma(\Theta) = & {\textstyle\sum\limits_n} \biggl[
\big(2 n + 1\big) \big(1-\cos\theta_n \big) - \beta^\gamma
 \frac12 \frac{1-\cos\big(2\theta_n\big) }{ (2n+1)^\gamma}\biggr]
  \\ \notag \hspace{-.5truecm}
 & +\beta^\gamma 
\frac12 {\textstyle\sum\!\sum\limits_{\hskip-0.4truecm n\neq m}}\biggl[\frac{1-\cos\big(\theta_n-\theta_m\big) }{ |n-m|^\gamma}
     - \frac{1-\cos\big(\theta_n+\theta_m\big) }{ (n+m+1)^\gamma}\biggr]
\end{align}
where the summations run over ${\NN}_0$; here, {$\Theta:=(\theta_n)^{}_{n\in{\NN}_0}$}, and $\beta=\frac{1}{2\pi T}$.
 The variations of 
$K_\gamma(\Theta)$ w.r.t. $\Theta$ are unconstrained, but now one obtains \textsl{non-linear} Euler--Lagrange equations
for any stationary point $\Theta^s$ of $K_\gamma(\Theta)$; viz., $\forall n\in{\NN}_0$:
\vspace{-.25truecm}
\begin{equation}\label{eq:EL}
\hspace{-.5truecm}
\big( 2n + 1 \big) \sin\theta_n^s + \beta^\gamma{\textstyle\sum\limits_{m \geq 0}}
\biggl[\frac{\sin\big(\theta_n^s-\theta_m^s\big) }{|n-m|^\gamma}
- \frac{\sin\big(\theta_n^s+\theta_m^s\big)}{(n+m+1)^\gamma }\biggr]=0,
\end{equation}
\textcolor{black}{with the stipulation that $\frac{\sin\left(\theta_n^s-\theta_n^s\right) }{|n-n|^\gamma} := 0$.}
 In the following we shall omit the superscript ${}^s$ from $\Theta^s$.

 The system of equations (\ref{eq:EL}) has infinitely many solutions when the $\theta_n$ are allowed to take
values in $[0,2\pi]$, restricted only by the asymptotic condition that $\theta_n\to 0$ rapidly enough when $n\to\infty$; 
see \cite{YuzAltPRB}. 
 However, we here are only interested in solutions that are putative minimizers of $H({\mathbf S}|{\mathbf N})$,
i.e. of $K_\gamma(\Theta)$.
 In \cite{YuzAltPRB} it was shown that a sequence $\Theta=(\theta_n)^{}_{n\in{\NN}_0}$ that minimizes 
$K_\gamma(\Theta)$, must have $\Theta\in [0,\frac\pi2]^{{\NN}_0} =:S$; i.e.,
all\footnote{Alternatively, all $\theta_n\in[-\frac\pi2,0]$; these choices are gauge equivalent.}
 $\theta_n\in[0,\frac\pi2]$.
 The normal state corresponds to the sequence of angles $\underline\Theta:=(\theta_n =0)^{}_{n\in{\NN}_0}$.
 This trivial solution of (\ref{eq:EL}) manifestly exists for all $\gamma>0$ and $T >0$.
 Next we inquire into the question of its linear stability versus its instability against modes $\Theta\in S$ for which
$K_\gamma(\Theta)$ converges absolutely.
\vspace{-5pt}

\section{Linear stability analysis of $\underline\Theta$}\label{sec:STAB}

 In this section we will show that for all $\gamma>0$ there exists a unique critical temperature $T_c(\gamma)>0$ 
such that the trivial solution $\underline\Theta$ is linearly stable for $T>T_c$, yet it is unstable against superconducting
perturbations for $T<T_c$. 
 Moreover, we will establish a novel variational principle that directly characterizes $T_c(\gamma)$.

 We note that $K_\gamma(\underline\Theta)= H(\mathbf{N}|\mathbf{N}) = 0$. 
 Now expanding $K_\gamma(\Theta)$ about $\Theta=\underline\Theta$ to second order in $\Theta$ yields the quadratic form
\begin{align}\label{eq:KtoSECndORDER}
\hspace{-.5truecm}
K_\gamma^{(2)}
(\Theta) = & {\sum\limits_n} \biggl[
 \frac{1}{2}\big(2 n + 1\big) \theta_n^2
-  \beta^\gamma\frac{\theta_n^2 }{ (2n+1)^\gamma}\biggr]
  \\
\notag &
+ \beta^\gamma\frac{1}{4}
{\sum\!\sum\limits_{\hskip-0.5truecm n\neq m}}
\biggl[\frac{\big(\theta_n-\theta_m\big)^2}{  |n-m|^\gamma}
- \frac{\big(\theta_n+\theta_m\big)^2 }{ (n+m+1)^\gamma}\biggr],
\end{align}
which for all $\gamma>0$ and $T>0$ is well-defined on the Hilbert space $\mathcal{H}$ of
sequences that satisfy $\|\Theta\|_{\mathcal{H}}^2:=\sum_{n\geq 0} (2n+1)\theta_n^2 < \infty$.
 If $K_\gamma^{(2)}(\Theta) \geq 0$ for all $\Theta\in\mathcal{H}$, with ``$=0$'' iff $\Theta = \underline\Theta$, 
then $K_\gamma(\Theta)>0$ for all $\Theta\neq\underline\Theta$ in a sufficiently small neighborhood of
$\underline\Theta$, which means that the trivial sequence $\underline\Theta$ is a local minimizer of $K_\gamma(\Theta)$
and thus linearly stable, then.
 If on the other hand there is at least one $\Theta\neq \underline\Theta$ in $\mathcal{H}\cap S$
for which $K_\gamma^{(2)}(\Theta)<0$, then the trivial sequence $\underline\Theta$ is 
{not a local minimizer} of $K_\gamma(\Theta)$ in $\mathcal{H}\cap S$, and therefore unstable against ``superconducting perturbations.''
 The verdict as to linear stability versus instability depends on $\gamma$ and $T$.

 To investigate this stability problem we first simplify $K_\gamma^{(2)}(\Theta)$. 
 After using the binomial formula $(\theta_n\pm\theta_m)^2 = \theta_n^2 \pm 2\theta_n\theta_m + \theta_m^2$, 
the contributions in the double sum that involve the $\theta_n^2$ and $\theta_m^2$ terms can be summed 
over the remaining index by noting that these sums of the kernel differences are telescoping.
 This allows us to separate the positive and negative contributions, viz.
\begin{align}\label{eq:K2simplified}
\hspace{-0.8truecm}
K_\gamma^{(2)}(\Theta) = &\, \frac12{\sum\limits_n} \biggl[
 \big(2 n + 1\big)  
             + \beta^\gamma {\sum\limits_{k=1}^{n}} \frac{2}{ k^\gamma} \biggr] \theta_n^2  \\
\notag &
- \beta^\gamma\frac{1}{2}
\sum\limits_n\!\sum\limits_m     
\theta_n\biggl[\frac{1 - \delta_{n,m}}{ |n-m|^\gamma} + \frac{1}{ (n+m+1)^\gamma}\biggr]\theta_m,
\end{align}
where in the double sum $n=m$ is now allowed, and where
it is understood that $\frac{1-\delta^{}_{n,n}}{ |n-n|^\gamma} := 0$.

 Next, we recast the functional $K_\gamma^{(2)}(\Theta)$ defined on $\mathcal{H}$ as a functional $Q_\gamma(\Xi)$ 
defined on $\ell^2({\NN}_0)$.
 For this we note that we can take the square root of the diagonal matrix $\mathfrak{O}$ whose diagonal elements are 
the odd natural numbers.
 Its square root is also a diagonal matrix, and its action on $\Theta$ componentwise is given as
\begin{equation}\label{eq:Dop}
(\mathfrak{O}^{\frac12}\Theta)_n =  \sqrt{2n + 1}\; \theta_n =: \xi_n.
\end{equation}
 Since $\Theta := (\theta_n)_{n\in\NN_0}\subset \mathcal{H}$, the sequence $\Xi:= (\xi_n)_{n\in\NN_0}\subset\ell^2(\NN_0)$.
The map $\mathfrak{O}^{\frac12}\!:\! \mathcal{H}\! \to\! \ell^2({\NN}_0)$ is invertible.
 Thus we set $K_\gamma^{(2)}(\Theta) =:\frac12 Q_\gamma(\Xi)$, viz.
 \begin{align}\label{eq:QofXI}
\hspace{-0.8truecm}
Q_\gamma(\Xi) = &\, {\sum\limits_n} \biggl[
1  + \beta^\gamma\frac{1}{2 n + 1} {\sum\limits_{k=1}^{n}} \frac{2}{ k^\gamma} \biggr] \xi_n^2  \\
\notag &
-\beta^\gamma 
\sum\!\sum\limits_{\hskip-0.5truecm n\neq m}
\xi_n  \biggl[\frac{1}{\sqrt{2 n + 1}}\,\frac{1}{ |n-m|^\gamma}\, \frac{1}{\sqrt{2 m + 1}}\biggr] \xi_m \\
\notag &
-\beta^\gamma 
\sum\limits_n\!\sum\limits_m     
\xi_n  \biggl[\frac{1}{\sqrt{2 n + 1}}\, \frac{1}{ (n+m+1)^\gamma}\, \frac{1}{\sqrt{2 m + 1}}\biggr] \xi_m.
\end{align}
\smallskip

We now state the main properties of the functional $Q_\gamma(\Xi)$, which decides the question of linear stability vs. 
instability of the normal state against superconducting perturbations.

\smallskip
\noindent
{\bf Theorem~1}: \textsl{For $\gamma>0$ and $T>0$, the functional $Q_\gamma$ given in (\ref{eq:QofXI}) has 
a minimum on the sphere $\big\{\Xi\in\ell^2(\NN_0): \|\Xi\|_{\ell^2}=1\big\}$.
 The minimizing (optimizing) eigenmode} $\Xi^{\mbox{\tiny{opt}}}$ \textsl{satisfies}
$\mathfrak{O}^{-\frac12}\Xi^{\mbox{\tiny{opt}}} \in [0,\frac\pi2]^{\mathbb{N}_0}$ \textsl{(after at most rescaling).
  Moreover, given $\gamma>0$ there is a unique $T_c(\gamma)>0$ at which
$\min\big\{Q_\gamma(\Xi) \!:\! \|\Xi\|_{\ell^2}=1\big\} =0$, and $\min\big\{Q_\gamma(\Xi) \!:\!\|\Xi\|_{\ell^2}=1\big\} >0$ 
when $T >T_c$, while $\min\big\{Q_\gamma(\Xi) \!:\! \|\Xi\|_{\ell^2}=1\big\} <0$ when $T <T_c$.
 Furthermore, the map $\gamma\mapsto T_c(\gamma)$ is continuous.}
\smallskip

 We prepare the proof of Theorem~1 by defining several linear operators that act on $\ell^2(\NN_0)$ and which are 
associated with $Q_\gamma$. 
 Letting $\big\langle \Xi,\widetilde\Xi\big\rangle$ denote the usual $\ell^2(\NN_0)$ inner product between
two $\ell^2$ sequences $\Xi$ and $\widetilde\Xi$, we write $Q_\gamma$ shorter thus:
 \begin{equation}\label{eq:QofXIrew}
Q_\gamma(\Xi) =  \bigl\langle \Xi\, , \big( \mathfrak{I} - \beta^\gamma\mathfrak{G}\big)  \Xi\bigr\rangle \,.
 \end{equation}
Here, $\mathfrak{I}$ is the identity operator, and $\mathfrak{G} =   - \mathfrak{G}_1 + \mathfrak{G}_2 + \mathfrak{G}_3$, where
the $\mathfrak{G}_j = \mathfrak{G}_j(\gamma)$ for $j\in\{1,2,3\}$ are Hilbert--Schmidt operators that map $\ell^2(\NN_0)$ 
compactly into $\ell^2(\NN_0)$, and that act as follows, componentwise:
 \begin{align}\label{eq:KopsComps}
\hspace{-0.8truecm}
(\mathfrak{G}_1(\gamma)\Xi)_n = &\, 
 \biggl[\frac{1}{2 n + 1} {\sum\limits_{k=1}^{n}} \frac{2}{k^\gamma}\biggr] \xi_n \,, \\
\hspace{-0.8truecm}
(\mathfrak{G}_2(\gamma)\Xi)_n = &\, 
\sum\limits_m \biggl[\frac{1}{\sqrt{2 n + 1}}\,\frac{1-\delta_{n,m}}{|n-m|^\gamma}\,\frac{1}{\sqrt{2 m+1}}\biggr]\xi_m\,,\\
\hspace{-0.8truecm}
(\mathfrak{G}_3(\gamma)\Xi)_n = &\, 
\sum\limits_m \biggl[\frac{1}{\sqrt{2 n + 1}}\, \frac{1}{ (n+m+1)^\gamma}\, \frac{1}{\sqrt{2 m + 1}}\biggr]\xi_m\,;
\end{align}
in appendix \ref{appendixA} we show that for $j\in\{1,2,3\}$, each $\mathfrak{G}_j\in \ell^2(\NN_0\times\NN_0)$ for all 
$\gamma>0$.
 Note also that $\mathfrak{G}_1$ is a diagonal operator with non-negative diagonal elements, 
$\mathfrak{G}_2$ is a real symmetric operator with vanishing diagonal elements and positive off-diagonal elements,
and $\mathfrak{G}_3$ is a real symmetric operator with all positive elements. 
 
 With the help of (\ref{eq:QofXIrew}) we will show that $Q_\gamma(\Xi)$ has a minimum on 
the sphere $\big\{\Xi\in\ell^2(\NN_0):  \|\Xi\|_{\ell^2}=1\big\}$, and that the minimizing $\ell^2(\NN_0)$ mode
$\Xi^{\mbox{\tiny{opt}}}$ satisfies the pertinent linear Euler--Lagrange equation
\begin{equation}\label{eq:ELinXIlin}
\big(  \mathfrak{I} -\beta^\gamma\mathfrak{G}(\gamma)\big)  \Xi = \kappa \Xi
\end{equation}
when $\kappa$ is the smallest eigenvalue of 
 $ \mathfrak{I} -\beta^\gamma\mathfrak{G}(\gamma)$.
 Since $\mathfrak{G}(\gamma)$ is independent of $T$, the boundary between linear stability and instability
of $\underline\Xi (=\underline\Theta)$, corresponding to $\kappa =0$, yields a unique $T_c (\gamma)$. 

 To establish all this we will take advantage of some well-known properties of compact operators that act on $\ell^2(\NN_0)$,
which for the convenience of the reader we collect in the following lemmas:

\smallskip
\noindent
{\bf Lemma~1}: \textsl{For $j\in\{1,2,...,N\}$, let $\mathfrak{C}_j:\ell^2(\NN_0)\to\ell^2(\NN_0)$
 be compact (and real symmetric), and let $a_j\in\RR$.
 Then also $\mathfrak{C} = \sum_j a_j\mathfrak{C}_j$ (is real symmetric and) maps $\ell^2(\NN_0)$ compactly into $\ell^2(\NN_0)$.}

\smallskip
\noindent
{\bf Lemma~2}: \textsl{Let $\mathfrak{C}:\ell^2(\NN_0)\to\ell^2(\NN_0)$ be compact, and let $\mathfrak{B}:\ell^2(\NN_0)\to\ell^2(\NN_0)$ be bounded.
 Then also $\mathfrak{B}\mathfrak{C}$ is compact.}

\smallskip
\noindent
{\bf Lemma~3}: \textsl{Let $\mathfrak{C}:\ell^2(\NN_0)\to\ell^2(\NN_0)$ be compact and real symmetric.
 Then its spectrum $\sigma(\mathfrak{C})$ is a bounded, countable subset of the real line. 
 The spectrum $\sigma(\mathfrak{C})$ contains $0$ either as eigenvalue of $\mathfrak{C}$ (of finite or possibly infinite 
multiplicity) or as accumulation point of $\sigma(\mathfrak{C})$, or possibly both.
 If $0$ is not an accumulation point of $\sigma(\mathfrak{C})$, then it is an eigenvalue with infinite multiplicity.
 All non-zero points in $\sigma(\mathfrak{C})$ are eigenvalues of $\mathfrak{C}$ of finite multiplicity.
}

\smallskip
\noindent
{\bf Lemma~4}: \textsl{Let $\mathfrak{C}:\ell^2(\NN_0)\to\ell^2(\NN_0)$ be compact and real symmetric, hence self-adjoint,
and let $\mathfrak{T}:\ell^2(\NN_0)\to \ell^2(\NN_0)$ be self-adjoint.
 Then also $\mathfrak{T}+\mathfrak{C}$ is self-adjoint, and its essential spectrum is that of $\mathfrak{T}$.}

\smallskip
\noindent
{\bf Lemma~5}: \textsl{Let $\mathfrak{C}:\ell^2(\NN_0)\to \ell^2(\NN_0)$ be compact and map the closed positive cone of $\ell^2(\NN_0)$ into itself.
  Then the spectral radius $\rho(\mathfrak{C})\in \sigma(\mathfrak{C})$. 
 Moreover, if $\rho(\mathfrak{C})>0$, then $\rho(\mathfrak{C})$ is a simple eigenvalue of $\mathfrak{C}$, and the 
pertinent real eigenmode can be chosen to have all its components positive (or negative, equivalently).
}

\smallskip
\noindent
{\bf Lemma~6}: \textsl{Let $\mathfrak{C}:\ell^2(\NN_0)\to\ell^2(\NN_0)$ be compact (and real symmetric).
 Then $\mathfrak{C}$ can be approximated arbitrarily precisely in operator norm by compact (real symmetric) operators of finite rank.}

\smallskip
\noindent
{\bf Remark~1}:
 \blue{Our Lemmas~1--6 are special cases of more general facts that can be found, for instance, in section VI of \cite{RSa} or section VI
of \cite{Brezis}, and earlier references therein.
 Note that Lemma~1 simply fleshes out what it means that the compact operators from $\ell^2(\NN_0)$ to $\ell^2(\NN_0)$ form
a Banach space, while Lemma~2 means that these compact operators form a left ideal in the Banach algebra of bounded operators
from $\ell^2(\NN_0)$ to $\ell^2(\NN_0)$.
 Lemma~3 is a special case of Theorem 6.8 of \cite{Brezis}.}
 Lemma~4 is known as Weyl's theorem, here specialized to real symmetric compact operators acting on the 
Hilbert space $\ell^2(\NN_0)$. 
 Lemma~5 is known as the Krein--Rutman theorem \blue{(Theorem 6.13 in \cite{Brezis})}, 
here specialized to compact operators acting on $\ell^2(\NN_0)$. 
 Lemma~6 allows us to practically work with an $N$-angle mode truncation of $\ell^2(\NN_0)$.
 In this case the Krein--Rutman theorem reduces to the \blue{more widely known} Perron--Frobenius theorem. 
\hfill $\square$

We are ready to prove Theorem~1.
 The proof will produce two useful characterizations of $T_c(\gamma)$ 
in terms of the spectrum of $\mathfrak{G}(\gamma)$.
 Understanding all quantities to depend on $\gamma$, we often omit the argument 
$\gamma$ from the ensuing formulas.

\smallskip
\noindent
\textsl{Proof of Theorem~1}: 
 We will show that $Q_\gamma(\Xi)$ is the quadratic form of the self-adjoint operator $\mathfrak{I} -\beta^\gamma\mathfrak{G}$. 
 We then show that $\mathfrak{I} -\beta^\gamma\mathfrak{G}$ has a smallest, 
non-degenerate eigenvalue that depends continuously and monotonically
on $T$, changing sign once when $T$ varies over the positive real line, with $\gamma$ fixed. 
 We also show that the components of the minimizing real eigenmode of $Q_\gamma(\Xi)$ do not change sign.

 In this vein, we begin by noting that all four operators, $\mathfrak{I}$, $\mathfrak{G}_1$, $\mathfrak{G}_2$, 
and $\mathfrak{G}_3$ are bounded and thus defined on all of $\ell^2(\NN_0)$; since they are real symmetric, they 
also are self-adjoint. 
 Since the operators $\mathfrak{G}_j$, $j\in\{1,2,3\}$, are Hilbert--Schmidt operators (see Appendix A), \blue{and since
Hilbert--Schmidt operators are compact (see \cite{Brezis}, \cite{RSa}),}
by Lemma~1 the operator $\mathfrak{G} = -\mathfrak{G}_1+\mathfrak{G}_2+\mathfrak{G}_3$ is compact, and 
since $\mathfrak{G}$ is also real-symmetric, it also is self-adjoint on $\ell^2(\NN_0)$. 
 By Lemma~4, the operator  $\mathfrak{I} -\beta^\gamma\mathfrak{G}$ is then a self-adjoint operator, too.
 Since therefore $Q_\gamma$ is the quadratic form of a self-adjoint operator, it suffices to discuss the spectrum of this 
self-adjoint operator.

 Recall, $\sigma\big(\mathfrak{I} - \beta^\gamma\mathfrak{G}\big) = 
\sigma_{\mbox{\tiny{ess}}}\big(\mathfrak{I}- \beta^\gamma\mathfrak{G}\big)\cup
\sigma_{\mbox{\tiny{disc}}}\big(\mathfrak{I}- \beta^\gamma\mathfrak{G}\big)$, 
a disjoint decomposition that is true anyhow for all operators; see section VII in \cite{RSa}.

 Now we first note that the essential spectrum of $\mathfrak{I}-\beta^\gamma\mathfrak{G}$
 consists of the single point $\{1\}$.
 This follows from Lemma~4, since obviously
$\sigma\big(\mathfrak{I}\big) =\sigma_{\mbox{\tiny{pp}}}\big(\mathfrak{I}\big) 
=\left\{1\right\} = \sigma_{\mbox{\tiny{ess}}}\big(\mathfrak{I}\big)$, while 
$\beta^\gamma\mathfrak{G}$ is compact by Lemma~2.

 Next we note that by Lemma~3, the non-essential spectrum of $\mathfrak{G}$ is bounded, countable, and consists 
of real eigenvalues of finite multiplicity, while the essential spectrum consists of the single point $\{0\}$. 
 Since any eigenmode $\Xi$ of $\mathfrak{G}$ is automatically also an eigenmode of the identity operator, 
the discrete spectrum of 
$\mathfrak{I}-\beta^\gamma\mathfrak{G}$ consists of the discrete spectrum of 
$-\beta^\gamma\mathfrak{G}$ shifted upward by $1$.
 And so, to show that 
$\mathfrak{I}-\beta^\gamma\mathfrak{G}$ has a smallest eigenvalue it suffices to show that $\mathfrak{G}$ 
has a largest eigenvalue.
 This in turn follows if we can show that $\sigma\big(\mathfrak{G}\big)$ contains a positive value, 
for then $\sigma\big(\mathfrak{G}\big)$ contains a largest positive value (because the spectrum of an operator is closed, 
and the spectrum of a compact operator is also bounded), and this largest positive value must be an eigenvalue of
$\mathfrak{G}$ because it lies above the essential spectrum $\{0\}$ of $\mathfrak{G}$.
 
 It is readily shown that $\mathfrak{G}$ has a positive eigenvalue by noting that $\langle\Xi,\mathfrak{G}\,\Xi\rangle$ is positive
 when $\Xi = \Xi_1 := (\xi_0,0,0,...)$ with $\xi_0\neq 0$.
Namely, since $\big\langle\Xi_1,\mathfrak{G}_1\Xi_1\big\rangle =0$ and $\big\langle\Xi_1,\mathfrak{G}_2\Xi_1\big\rangle =0$, 
too, while $\big\langle\Xi_1,\mathfrak{G}_3\Xi_1\big\rangle = \xi_0^2 >0$, 
we have $\big\langle\Xi_1,\mathfrak{G}\,\Xi_1\big\rangle = \big\langle\Xi_1,\mathfrak{G}_3\,\Xi_1\big\rangle >0$.
 The compactness of $\mathfrak{G}$ now implies that there is a normalized eigenmode $\Xi^{\mbox{\tiny{opt}}}$ that
maximizes $\big\langle\Xi,\mathfrak{G}\,\Xi\big\rangle$, 
and $\big\langle\Xi^{\mbox{\tiny{opt}}},\mathfrak{G}\,\Xi^{\mbox{\tiny{opt}}}\big\rangle >0$.
 This same mode $\Xi^{\mbox{\tiny{opt}}}$ then manifestly minimizes 
$\big\langle\Xi,\big(\mathfrak{I}-\beta^\gamma\mathfrak{G}\big)\Xi\big\rangle= Q_\gamma(\Xi)$ 
over the normalized sphere in $\ell^2(\NN_0)$.
 The suffix ``${}^{\mbox{\tiny{opt}}}$'' stands for ``optimizer,'' to avoid any confusion that could entail when
using the suffix ``${}^{\mbox{\tiny{max}}}$'' or ``${}^{\mbox{\tiny{min}}}$.'' 

 Next, since this optimizer $\Xi^{\mbox{\tiny{opt}}}$ only depends on $\gamma$, but not on $T$, it 
follows right away that when  $\gamma>0$ is fixed, then 
$Q_\gamma(\Xi^{\mbox{\tiny{opt}}}) =  \big\langle\Xi^{\mbox{\tiny{opt}}},\Xi^{\mbox{\tiny{opt}}}\big\rangle 
- \beta^\gamma\big\langle\Xi^{\mbox{\tiny{opt}}},\mathfrak{G}\,\Xi^{\mbox{\tiny{opt}}}\big\rangle$ is a continuous, monotone 
increasing function of $T$, which goes to $-\infty$ in the limit $T\searrow 0$, and is $=0$ iff $\beta^\gamma\mathfrak{g}(\gamma)=1$, 
where $\mathfrak{g}(\gamma)>0$ denotes the largest eigenvalue of $\mathfrak{G}(\gamma)$. 
 This gives us $T = T_c(\gamma)$ with
\begin{equation}\label{eq:TAUcVPkappa}
T_c(\gamma) =  \tfrac{1}{2\pi}\big[\mathfrak{g}(\gamma)\big]^{\frac1\gamma}.
\end{equation}

 We remark that  $\mathfrak{g}(\gamma)$ also is the spectral radius of $\mathfrak{G}(\gamma)$,
a useful fact that we prove later.

 Next, since all components of the operator $\mathfrak{G}(\gamma)$ depend continuously on $\gamma$,
then also the eigenvalues of $\mathfrak{G}(\gamma)$ depend continuously on $\gamma$, and therefore
so does $T_c(\gamma)$, as claimed.
 
 The remaining claim to be verified is that, after at most multiplication by a positive constant (because we face a linear
problem), $\Theta^{\mbox{\tiny{opt}}}\in \big(0,\frac\pi2\big]^{\NN_0}$ (up to an irrelevant overall sign change).
 This translates into the statement that (up to an irrelevant overall sign change)
 $\Xi^{\mbox{\tiny{opt}}}$ is in the closed positive cone of $\ell^2(\NN_0)$, denoted $\ell^2_{\geq 0}$.
 Recalling that $\mathfrak{G} = - \mathfrak{G}_1 + \mathfrak{G}_2 + \mathfrak{G}_3$,
we rewrite the eigenvalue equation 
\begin{equation}\label{eq:ELforXImin}
\mathfrak{G}   \Xi^{\mbox{\tiny{opt}}}  = \mathfrak{g}\, \Xi^{\mbox{\tiny{opt}}} 
\end{equation}
for the largest eigenvalue $\mathfrak{g}$ of $\mathfrak{G}$ into
\begin{equation}
\big(\mathfrak{G}_2  + \mathfrak{G}_3 \big) \Xi^{\mbox{\tiny{opt}}} 
= 
\big(\mathfrak{G}_1 + \mathfrak{g}   \mathfrak{I}\big)\Xi^{\mbox{\tiny{opt}}}.
\end{equation}
 Now recall that $\mathfrak{g} >0$, and that $\mathfrak{G}_1$ is
a non-negative diagonal operator.
 It follows that $\mathfrak{G}_1 + \mathfrak{g} \mathfrak{I}$ is a positive diagonal operator, hence invertible.
 And so $\Xi^{\mbox{\tiny{opt}}}$ (the eigenmode of the largest eigenvalue of $\mathfrak{G}$) satisfies
\begin{equation}\label{eq:BS}
\big(\mathfrak{G}_1 + \mathfrak{g} \mathfrak{I}\big)^{-1}\big(\mathfrak{G}_2  + \mathfrak{G}_3 \big) 
\Xi^{\mbox{\tiny{opt}}} = \Xi^{\mbox{\tiny{opt}}} .
\end{equation}
 Recall next that $\mathfrak{G}_1$ is a compact operator.
 Thus $\mathfrak{G}_1 + \mathfrak{g} \mathfrak{I}$ has a spectrum that is a bounded, countable subset of the real line. 
 Since $\mathfrak{G}_1$ is non-negative, and since $\mathfrak{g}>0$, it follows that the spectrum of 
$\mathfrak{G}_1 + \mathfrak{g} \mathfrak{I}$ is bounded away from 0 by $\mathfrak{g} >0$. 
 Hence also $\big(\mathfrak{G}_1 + \mathfrak{g} \mathfrak{I}\big)^{-1}$ is a positive, diagonal, and bounded operator
(its spectrum is bounded from above by $\frac{1}{\mathfrak{g}}$).
 Recall next that the $\mathfrak{G}_j$ are compact, real-symmetric operators.
 By Lemma~1 and Lemma~2 in concert, 
$\big(\mathfrak{G}_1 + \mathfrak{g} \mathfrak{I}\big)^{-1}\big(\mathfrak{G}_2  + \mathfrak{G}_3 \big) =: \mathfrak{C}(\mathfrak{g})$
is a compact operator.
 Moreover, since each $\mathfrak{G}_j$ with $j\in\{1,2,3\}$ maps the closed positive cone $\ell^2_{\geq 0}(\NN_0)$ into itself, 
and since so does $\mathfrak{I}$, we conclude that  the compact operator $\mathfrak{C}(\mathfrak{g})$
maps $\ell^2_{\geq 0}(\NN_0)$ into itself, too.
 Since it is not the zero operator, Lemma~5 now applies and guarantees that its spectral radius is the largest eigenvalue,
and that the pertinent eigenmode has only non-zero elements of the same sign --- which can be chosen to be positive.

 We now show that the spectral radius of $\mathfrak{C}(\mathfrak{g})$ equals 1.
 For then its pertinent eigenmode is $\Xi^{\mbox{\tiny{opt}}}$, and as such has only positive elements 
(after choosing an overall sign).

 To see that 1 is the spectral radius of $\mathfrak{C}(\mathfrak{g})$, note first that (\ref{eq:BS}) reveals that 1 is 
in the spectrum of $\mathfrak{C}(\mathfrak{g})$.
 Therefore its spectral radius $\rho\big(\mathfrak{C}(\mathfrak{g})\big)\geq 1$. 
 Now suppose that  $\rho\big(\mathfrak{C}(\mathfrak{g})\big)> 1$. 
 Replacing the fixed eigenvalue $\mathfrak{g}  >0$ by a parameter $\varkappa>0$, we note that
the operator $\mathfrak{C}(\varkappa)$ depends continuously on $\varkappa$.
 In fact, $\mathfrak{C}(\varkappa)$ decreases monotonically when $\varkappa$ increases; in particular, it decreases to the 
zero operator when $\varkappa\to \infty$. 
 Therefore, since by hypothesis $\rho\big(\mathfrak{C}(\varkappa)\big)>1$ when $\varkappa=\mathfrak{g}$, we can increase $\varkappa$ 
and thereby continuously lower the spectral radius $\rho\big(\mathfrak{C}(\varkappa)\big)$ until it equals 1, at (say)
 $\varkappa=\varkappa_* > \mathfrak{g} $. 
 But then $\mathfrak{C}(\varkappa_*) \Xi_* = \Xi_*$ for the pertinent eigenmode $\Xi_*$ of the eigenvalue 
$1=\rho\big(\mathfrak{C}(\varkappa_*)\big)$, 
and by reversing the steps that lead from (\ref{eq:ELforXImin}) to (\ref{eq:BS}), we see that $\Xi_*$ is also 
an eigenmode of $\mathfrak{G}$, though with eigenvalue $\varkappa_* > \mathfrak{g}$. 
 This contradicts the fact that $\mathfrak{g} >0$ is by definition the largest eigenvalue of $\mathfrak{G}$.
 Hence $\rho\big(\mathfrak{C}(\mathfrak{g})\big) = 1$.
\hfill {\textbf{Q.E.D.}}
\smallskip

 As announced in the proof of Theorem~1, it is useful to add the following non-obvious fact about the spectrum of 
the operator $\mathfrak{G}(\gamma)$.

\smallskip
\noindent
{\bf Proposition~1}: 
\textsl{Let $\gamma>0$ be given. Then the largest eigenvalue $\mathfrak{g}(\gamma)$ of
$\mathfrak{G}(\gamma)$ is also the spectral radius $\rho\big(\mathfrak{G}(\gamma)\big)$.}

\smallskip
\noindent
\textsl{Proof of Proposition~1}: 
 Suppose that $\mathfrak{g}\neq \rho(\mathfrak{G})$. 
 Since $\mathfrak{g}>0$ is the largest eigenvalue of $\mathfrak{G}$, there then 
is a negative eigenvalue $\kappa<0$ of $\mathfrak{G}$ with magnitude $|\kappa| >\mathfrak{g}$,
and $\rho(\mathfrak{G}) = - \kappa$.
 By the compactness of $\mathfrak{G}$, the eigenspace of $\kappa$ is finite-dimensional.
 Let $\Xi_-$ denoted a normalized eigenmode in this eigenspace.

 Since $\kappa = \min\sigma(\mathfrak{G})$, the operator $\mathfrak{G} - \kappa\mathfrak{I}$ is non-negative, with
essential spectrum $\{-\kappa\}$, and with a non-trivial kernel space $\mathcal{N}:=$ ker$(\mathfrak{G} - \kappa\mathfrak{I})$; i.e.,
$\min_\Xi \langle \Xi, (\mathfrak{G} - \kappa\mathfrak{I})\Xi\rangle = 0$ (minimization over normalized $\Xi$ is understood),
and $\mathcal{N}$ is finite-dimensional.
 Furthermore, since $\mathfrak{G} = -\mathfrak{G}_1 + \mathfrak{G}_2 + \mathfrak{G}_3$, and all $\mathfrak{G}_j$ 
are non-negative, the operator $\mathfrak{G}_2 + \mathfrak{G}_3 - \kappa\mathfrak{I}$ is non-negative, too.

 In fact, $\mathfrak{G}_2 + \mathfrak{G}_3 - \kappa\mathfrak{I}$ is positive; 
for assume first that $\Xi\not\in \mathcal{N}$, then we have 
$\inf_{\Xi\in\mathcal{N}^\perp} \langle \Xi, (\mathfrak{G} - \kappa\mathfrak{I})\Xi\rangle = B >0$ because $0$ is an
isolated eigenvalue of $\mathfrak{G} - \kappa\mathfrak{I}$, and then
also $\inf_{\Xi\in\mathcal{N}^\perp} \langle \Xi, (\mathfrak{G}_2+\mathfrak{G}_3 - \kappa\mathfrak{I})\Xi\rangle = \tilde{B} >0$,
with $\tilde{B}\geq B$, by the non-negativity of $\mathfrak{G}_1$. 
 Assume next that $\Xi\in \mathcal{N}$, i.e. $\Xi=\Xi_-$. 
 But then $\langle \Xi_-, (\mathfrak{G}_2+\mathfrak{G}_3 - \kappa\mathfrak{I})\Xi_-\rangle >0$; for $\Xi_-\in\mathcal{N}$ 
together with $(\mathfrak{G}_2+\mathfrak{G}_3 - \kappa\mathfrak{I})\Xi_- = 0$ would imply that
also $\mathfrak{G}_1\Xi_-=0$; but, this is impossible because $\mathfrak{G}_1$ is a non-negative diagonal operator 
with one-dimensional kernel space spanned by $\Xi_1 =(\xi_0,0,\cdots)$ for some $\xi_0\neq 0$, and $\Xi_1\not\in\mathcal{N}$
since
$(\mathfrak{G}_2 + \mathfrak{G}_3 - \kappa\mathfrak{I}) \Xi_1 = \big(1-\kappa\big)\Xi_1$, with
$-\kappa>0$.
 It therefore follows that $\mathfrak{G}_2 + \mathfrak{G}_3 - \kappa\mathfrak{I}$ is positive.

 And so, since $\mathfrak{G}_2 +\mathfrak{G}_3 - \kappa\mathfrak{I}$ is self-adjoint, positive, and bounded away from zero,
it is invertible with a bounded inverse.
 Thus we can rewrite the eigenvalue problem $\mathfrak{G}\Xi_- = \kappa\Xi_-$ as 
\begin{equation}\label{eq:XiMINUSev}
\left(\mathfrak{G}_2 + \mathfrak{G}_3 -\kappa\mathfrak{I}\right)^{-1} \mathfrak{G}_1\Xi_- = \Xi_-.
\end{equation}
 By an obvious variation of the reasoning in the last part of the proof of Theorem~1, we now conclude that
$1$ is the spectral radius of the compact, positive operator
$\left(\mathfrak{G}_2 + \mathfrak{G}_3 -\kappa\mathfrak{I}\right)^{-1} \mathfrak{G}_1$. 
 The Krein--Rutman theorem now implies that the associated eigenspace is one-dimensional, and is spanned by
a real eigenmode, $\Xi_-$, whose components are all positive (after choosing an overall sign).
 This implies that $\langle \Xi_-,\Xi^{\mbox{\tiny{opt}}}\rangle >0$; but this is impossible, becauses the 
eigenmodes of a self-adjoint operator that belong to different eigenvalues are orthogonal. 
 Therefore, $\rho(\mathfrak{G}) = \mathfrak{g}$.
\hfill {\textbf{Q.E.D.}}
\smallskip

 Proposition~1 allows us to characterize the critical $T_c$ as follows:
\begin{equation}\label{eq:TcVPrho}
T_c(\gamma) =  \frac{1}{2\pi}\big[\rho\big(\mathfrak{G}(\gamma)\big)\big]^{\frac1\gamma} .
\end{equation}

\section{Rigorous lower bounds on~$T_c(\gamma)$}\label{sec:lowerB}

 The explit determination of $\mathfrak{g} (\gamma)$ in closed form may not be feasible.
 Yet recall that (\ref{eq:TAUcVPkappa}) poses a \textsl{variational principle} directly for $T_c$, viz.
\begin{equation}\label{eq:TAUcVP}
T_c(\gamma) : = \frac{1}{2\pi} \biggl(\max_\Xi\frac{\big\langle\Xi,\mathfrak{G}(\gamma) \,\Xi\big\rangle}
        {\big\langle\Xi,\Xi\big\rangle}\biggr)^{\frac1\gamma} ,
\end{equation}
where the maximum is taken over non-vanishing $\Xi\in\ell^2(\NN_0)$.
 Having the variational principle (\ref{eq:TAUcVP}) for $T_c(\gamma)$, we in principle can get
arbitrarily accurate lower approximations to it by working with cleverly chosen trial sequences $\Xi$ in (\ref{eq:TAUcVP}).

 Rigorous lower bounds on $T_c(\gamma)$
are  obtained by truncation of $\ell^2(\NN_0)$ to $N$-angle trial sequences 
$\Xi_N:= (\xi_0,\xi_1,\dots,\xi_{N-1},0,0,\dots)$, with $\xi_j>0$ for $j\in\{0,...,N-1\}$ and $N\in\NN$.
 Evaluating (\ref{eq:TAUcVP}) with $\Xi_N$ in place of $\Xi^{\mbox{\tiny{opt}}}$ yields a monotonically increasing
sequence of lower bounds on ${T}_c(\gamma)$, viz.
\begin{equation}\label{eq:TAUcVPtrialN}
{T}_c^{(N)}(\gamma) : = 
\frac{1}{2\pi}\left(\max_{\Xi_N} \frac{\big\langle\Xi_N,\mathfrak{G}(\gamma)\,\Xi_N\big\rangle}
        {\big\langle\Xi_N,\Xi_N\big\rangle}\right)^{\frac1\gamma}.
\end{equation}

 The evaluation of (\ref{eq:TAUcVPtrialN}) is equivalent to solving the eigenvalue problem for the truncated variational 
principle, a finite-dimensional real symmetric matrix problem.
 With the help of algebraic software like Maple or Mathematica this can be accomplished in closed form when $N\in\{1,2,3,4\}$;
yet these expressions get unwieldy soon.
 For $N>4$ a numerical evaluation is necessary, requiring choices of $\gamma$.

The next theorem states that our lower approximations to $T_c(\gamma)$ as function of $\gamma$ can be made arbitrarily precise.

 \smallskip
\noindent
{\bf Theorem~2}: \textsl{Let $\gamma>0$ be given. 
 Let $\mathfrak{g}^{(N)}(\gamma)$ denote the largest eigenvalue of $\mathfrak{G}(\gamma)$ 
on the subset of normalized $N$-angle sequences $\Xi_N$. 
 Then ${T}_c^{(N)}(\gamma)$ defined by (\ref{eq:TAUcVPtrialN}) reads
\begin{align}\label{eq:TcVPapproxN}
 {T}_c^{(N)}(\gamma) =
\, \tfrac{1}{2\pi} \bigl[\mathfrak{g} ^{(N)}\bigl(\gamma\bigr) \bigr]^{\frac1\gamma}.
\end{align}
 The expressions ${T}_c^{(N)}(\gamma)$ given by (\ref{eq:TcVPapproxN}) converge upward to the
continuous function ${T}_c(\gamma)$ when $N\to\infty$.}

 \smallskip
\noindent
\textsl{Proof}: 
By Lemma~6, the maximizing sequences $\Xi_N$ for (\ref{eq:TAUcVPtrialN}) 
produce an increasing sequence of lower approximations $\mathfrak{g} ^{(N)}$ to $\mathfrak{g}$ that converges to $\mathfrak{g}$.  
 Evidently then (\ref{eq:TcVPapproxN}) is an increasing 
sequence of ${T}_c^{(N)}$ that converges upward to ${T}_c$, which is
continuous in $\gamma$ because $\mathfrak{g}$ is.  
\hfill {\textbf{Q.E.D.}}
\smallskip

 When $N\leq 4$ then ${T}_c^{(N)}(\gamma)$ can be worked out explicitly, given $\gamma >0$.
 This follows from the following known results about the characteristic polynomial of real symmetric $N\times N$ matrices,
here specialized for our needs.

\smallskip
\noindent
{\bf Lemma~7}: \textsl{Let $\mathfrak{M}:\RR^N \to\RR^N$ be a real symmetric $N\times N$ matrix, and $\mathcal{I}$ the
corresponding identity matrix. 
 Then the coefficients $c_k$ of its characteristic polynomial 
$\det\big(\mu\mathcal{I}-\mathfrak{M}\big) =: \sum_{k=0}^N c_k\mu^k$ 
are themselves explicitly known polynomials of degree $N-k$ in $\mathrm{tr}\, \mathfrak{M}^j$, $j\in\{1,...,N\}$.}
\smallskip

\noindent
{\bf Corollary~1}:
 \textsl{For $N\leq 4$ the spectrum $\sigma(\mathfrak{M})$ consists of explicit algebraic expressions involving
these polynomials in $\mathrm{tr}\, \mathfrak{M}^j$, $j\in\{1,...,N\}$.}
\smallskip

 We illustrate Corollary~1 for the cases $N\in\{2,3,4\}$, skipping the case $N=1$ here because it is trivial.
 In our illustrations we refrain from unfolding some standard invariants such as $\det \mathfrak{M}$ into
their equivalent polynomials in  $\mathrm{tr}\, \mathfrak{M}^j$, $j\in\{1,...,N\}$. 
 In this vein, for the cases $N\in\{2,3,4\}$ one has the following.
\smallskip

When $N=2$, the familiar ``quadratic formula'' gives
\begin{align}\label{eq:2x2spec}
\sigma(\mathfrak{M}) 
=\Bigl\{\tfrac12\Big(
{\rm tr}\,\mathfrak{M} + j \sqrt{\big({\rm tr}\,\mathfrak{M}\big)^2 -4 \det\mathfrak{M}}\Big)\Bigr\}_{j\in\{\pm 1\}}^{}.
\end{align}
\smallskip

When $N=3$, Vi\'ete's formula yields
\begin{align}\label{eq:3x3spec}
&\hspace{-0.5truecm}
\sigma(\mathfrak{M}) = \\
\notag
&\hspace{-0.5truecm}
\left\{\!
\textstyle{\frac13 {\rm tr}\,\mathfrak{M} +  
2 \sqrt{\frac{p}{3}}\cos\! \left[\frac13\arccos\!\left(\!\frac{q}{2}\sqrt{\!\big(\frac3p\big)^{\!3}}\right)\!
-j\frac{2\pi}{3}\right] }\! \right\}_{j\in\{0,1,2\}}^{},
\end{align}
with 
\begin{align}\label{eq:3x3specP}
p = \tfrac13\big({\rm tr}\,\mathfrak{M}\big)^2- {\rm tr\, adj}\, \mathfrak{M}
\end{align}
and
\begin{align}\label{eq:3x3specQ}
q = 
 \tfrac{2}{27}\big({\rm tr}\,\mathfrak{M}\big)^3 
- \tfrac13 \big({\rm tr}\,\mathfrak{M}\big) \big({\rm tr\, adj}\,\mathfrak{M}\big)
+ \det\mathfrak{M},
\end{align}
where adj$\,\mathfrak{M}$ denotes the \textsl{adjugate matrix} to $\mathfrak{M}$.

When $N=4$, Cardano's formula yields (cf. \cite{Krvavica})
\begin{align}\label{eq:4x4spec}
&\hspace{-0.75truecm}
\sigma(\mathfrak{M}) = \\
\notag
&\hspace{-0.75truecm}
\left\{\! \!-\tfrac14 A + j\left[ \sqrt{\tfrac12 Z}
+k\!\sqrt{\!\tfrac{3}{16} A^2 -\tfrac12 B - \tfrac12 Z -j\tfrac{A^3 -4AB + 8C}{16\sqrt{2Z}}}\right]\right\}_{(j,k)\in\{\pm1\}^2}^{}
\end{align}
where $Z$ is a positive root of the associated resolvent cubic, which in this 
all-real-roots situation is conveniently given by Vi\'ete's formula
\begin{align}
\label{eq:ZdefAGAIN}
Z = \tfrac13 \Big[ \sqrt{Y} \cos\Big(\tfrac13 \arccos \tfrac{X}{2\sqrt{Y^3}}\Big) - B + \tfrac38 A^2\Big],
\end{align}
with  
\begin{align}\label{eq:XdefAGAIN}
X &= 2B^3 - 9ABC+27C^2 + 27A^2D -72BD,\\
Y &= B^2-3AC+12D,\label{eq:YdefAGAIN}
\end{align}
where
\begin{align}\label{eq:AdefM}
A &= - {\rm tr}\, \mathfrak{M}\\
\label{eq:BdefM}
B &= \tfrac12\Big( \big({\rm tr}\, \mathfrak{M}\big)^2 - {\rm tr}\, {\mathfrak{M}}^2\Big),\\
\label{eq:CdefM}
C &= -\tfrac16 \Big( 
   \big({\rm tr}\, \mathfrak{M}\big)^3 - 3 \big({\rm tr}\, {\mathfrak{M}}^2\big) \big({\rm tr}\, \mathfrak{M}\big)
        + 2{\rm tr}\, {\mathfrak{M}}^3\Big),\\
\label{eq:DdefM}
D &= \det {\mathfrak{M}}.
\end{align}

 With the help of these spectral formulas, and the trivial case when $N=1$, 
we now work out the lower bounds on ${T}_c(\gamma)$ explicitly for when $N\in\{1,2,3,4\}$. 
 In the following, we let $\mathfrak{G}^{(N)}$ denote the upper left $N\times N$ block of $\mathfrak{G}$. 
 Its spectrum is manifestly identical to the spectrum of the restriction of $\mathfrak{G}$ to the $N$-dimensional
subspace of $\ell^2(\NN_0)$ that consists of sequences $\Xi_N = (\xi_0,\xi_1,...,\xi_{N-1},0,0,...)$. 
 In particular, $\mathfrak{g}^{(N)}$ is the largest eigenvalue of $\mathfrak{G}^{(N)}$.

\subsection{The lower bound ${T}_c^{(1)}(\gamma)$} 
 
 We already invoked the single-angle truncation $\Xi_1 := (\xi_0,0,0,...)$ of $\ell^2(\NN_0)$
to show that $\mathfrak{G}$ has a positive eigenvalue on $\ell^2(\NN_0)$, bounded below by 
\begin{equation}\label{eq:kONE}
\mathfrak{g}^{(1)}(\gamma)= 1,
\end{equation}
which is the one and only eigenvalue of $\mathfrak{G}^{(1)}(\gamma) = (1)$.
 Inserted in (\ref{eq:TcVPapproxN}), this leads straight to the lower bound ${T}_c^{}(\gamma)\geq{T}_c^{(1)}(\gamma)$, 
with
\begin{align}\label{eq:TcVPapproxUNO}
T_c^{(1)}(\gamma) = \tfrac{1}{2\pi}.
\end{align}
{Replacing $T_c^{(1)}(\gamma)\to T_c^{(1)}(g,\gamma)/g$} yields (\ref{eq:TcLOWERboundsONE}).

\subsection{The lower bound ${T}_c^{(2)}(\gamma)$}

 The two-angle truncation yields the $2\times2$ matrix
\begin{equation}\label{eq:Ktwo}
 \mathfrak{G}^{(2)} = 
\begin{pmatrix}
   1 
& \frac{1}{2^\gamma \sqrt{3}} + \frac{1}{\sqrt{3}} \\
  \frac{1}{2^\gamma \sqrt{3}} + \frac{1}{\sqrt{3}} 
&  \frac{1}{3^{\gamma +1}} - \frac{2}{3} \\ 
\end{pmatrix}\!.
\end{equation}
  The invariants ${\rm tr}\,\mathfrak{G}^{(2)}$ and $\det\mathfrak{G}^{(2)}$ of a $2\times2$ matrix are readily computed as
\begin{equation}\label{eq:traceKtwo}
{\rm tr}\,\mathfrak{G}^{(2)} = \tfrac13 \big( 1 + \!\tfrac{1}{3^\gamma} \big)
\end{equation}
and
\begin{equation}\label{eq:determinantKtwo}
\det\mathfrak{G}^{(2)} = - \tfrac13 
\left(\!\big(1 + \tfrac{1}{2^\gamma}\big)^2\!\! +\! 2 - \tfrac{1}{3^\gamma} \right).
\end{equation}
 Equation (\ref{eq:traceKtwo}) reveals that ${\rm tr}\,\mathfrak{G}^{(2)}>0$, 
and (\ref{eq:determinantKtwo}) reveals that $\det\mathfrak{G}^{(2)}<0$.
 Hence, $\mathfrak{g}^{(2)}$ is obtained by setting $\mathfrak{M} = \mathfrak{G}^{(2)}$
in (\ref{eq:2x2spec}) and choosing the eigenvalue with $j=1$.
 And so, (\ref{eq:TcVPapproxN}) yields the lower bound ${T}_c^{}(\gamma) \geq {T}_c^{(2)}(\gamma)$, with
\begin{equation}\label{eq:hatTcTWO}
{T}_c^{(2)}(\gamma) = 
\frac{1}{2\pi}\left( \tfrac12\Big(
{\rm tr}\,\mathfrak{G}^{(2)}+\sqrt{\big({\rm tr}\,\mathfrak{G}^{(2)}\big)^2-4\det\mathfrak{G}^{(2)}}\,\Big)\!\!\right)^\frac1\gamma,
\end{equation}
where the trace and determinant of $\mathfrak{G}^{(2)}$ are given in (\ref{eq:traceKtwo}) and (\ref{eq:determinantKtwo});
explicitly:
\begin{equation}\label{eq:TcVPapproxDUEexplicit}
\hspace{-1truecm}
 T_c^{(2)}(\gamma)  =  \tfrac{1}{2\pi}\!
\left[\tfrac16\!\left(\!\! 1 \!+\!\tfrac{1}{3^\gamma} \!+\! 
\sqrt{\!\big(1 + \tfrac{1}{3^\gamma}\big)^2\! + \! 12\left(\!\! \big(1 + \tfrac{1}{2^\gamma}\big)^2\!\! 
+\! 2 - \tfrac{1}{3^\gamma} \right)\!}\right)\!\right]^{\!\frac1\gamma}\!\!\!\!.\!\! 
\end{equation}
 Note that 
r.h.s.(\ref{eq:TcVPapproxDUEexplicit})$\sim \frac{1}{2\pi}\left(\frac16\left[1+\sqrt{37}\right]\right)^{\!\frac1\gamma}\!$ 
as $\gamma\to\infty$. 
 Note furthermore that r.h.s.(\ref{eq:TcVPapproxDUEexplicit})$\sim\frac{1}{2\pi}\left(\frac53\right)^{\!\frac1\gamma}\!$ when $\gamma\searrow 0$.
 Hence, $T_c^{(2)}(\gamma) \searrow\frac{1}{2\pi}$ as $\gamma\to\infty$, but it 
blows up to $\infty$ faster than any inverse power of $\gamma$ when $\gamma\searrow 0$.
{Replacing $T_c^{(2)}(\gamma)\to T_c^{(2)}(g,\gamma)/g$} yields (\ref{eq:TcLOWERboundsTWO}).

\subsection{The lower bound ${T}_c^{(3)}(\gamma)$}

 The three-angle truncation yields the $3\times3$ matrix
\begin{equation}\label{eq:Kthree}
\mathfrak{G}^{(3)}(\gamma) = 
 \begin{pmatrix}
   {1}
&  \frac{1}{\sqrt{3}}\left(\frac{1}{2^\gamma }  + 1\right)
&  \frac{1}{\sqrt{5}}\left(\frac{1}{3^\gamma }  + \frac{1}{2^\gamma } \right) \\
   \frac{1}{\sqrt{3}}\left(\frac{1}{2^\gamma }  + 1\right)
&  \frac{1}{3}\left(\frac{1}{3^\gamma }  -2\right)
&  \frac{1}{\sqrt{15}}\left(\frac{1}{4^\gamma }  + 1\right) \\ 
   \frac{1}{\sqrt{5}}\left(\frac{1}{3^\gamma }  + \frac{1}{2^\gamma } \right) 
&  \frac{1}{\sqrt{15}}\left(\frac{1}{4^\gamma }  + 1\right) 
&  \frac{1}{5}\left(\frac{1}{5^\gamma} - \frac{2}{2^\gamma }  -2 \right)   \\
\end{pmatrix}\!.
\end{equation} 
 With the help of Maple we found that $\mathfrak{G}^{(3)}(\gamma)$ has exactly one positive and two negative eigenvalues.
 The largest eigenvalue $\mathfrak{g}^{(3)}(\gamma)>0$ is obtained by 
setting $\mathfrak{M} = \mathfrak{G}^{(3)}(\gamma)$ in (\ref{eq:3x3spec})--(\ref{eq:3x3specQ}) 
and choosing the eigenvalue with $j=0$.
 Since we already showed that the largest eigenvalue $\mathfrak{g}^{(N)}(\gamma)$ 
of $\mathfrak{G}^{(N)}(\gamma)$ maps into the largest eigenvalue of an $N\times N$ matrix with all positive entries, 
the Perron--Frobenius theorem applies and guarantees that $\mathfrak{g}^{(N)}(\gamma)$ is non-degenerate. 
\begin{align}\label{eq:3x3k}
\hspace{-0.7truecm}
{T}_c^{(3)}(\gamma) = &
\tfrac{1}{2\pi}\left(\! \textstyle{\frac13 {\rm tr}\,\mathfrak{G}^{(3)} +
2 \sqrt{\frac{p}{3}}\cos \left[\frac13\arccos\left(\frac{q}{2}\sqrt{\!\Big(\frac3p\Big)^{\!{}_3}}\,\right)\!\right] 
} \right)^\frac1\gamma\!\!\!,\!\!\\
\label{eq:3x3kP}
p = & \tfrac13\big({\rm tr}\,\mathfrak{G}^{(3)}\big)^2- {\rm tr\, adj}\, \mathfrak{G}^{(3)},\\
\label{eq:3x3kQ}
q = &
 \tfrac{2}{27}\big({\rm tr}\,\mathfrak{G}^{(3)}\big)^3 
- \tfrac13 \big({\rm tr}\,\mathfrak{G}^{(3)}\big) \big({\rm tr\, adj}\,\mathfrak{G}^{(3)}\big)
+ \det\mathfrak{G}^{(3)},
\end{align}
where $\mathfrak{G}^{(3)}$ is given in (\ref{eq:Kthree}).
 Also the invariants ${\rm tr}\,\mathfrak{G}^{(3)}$, ${\rm tr\, adj}\,\mathfrak{G}^{(3)}$, 
and $\det\mathfrak{G}^{(3)}$ of a $3\times3$ matrix are readily computed. 
  For the sake of completeness we supply these expressions explicitly in Appendix~B.2.

 We remark that for $\gamma\to\infty$, we have the asymptotic behavior
$T_c^{(3)}(\gamma) \sim\!\frac{1}{2\pi}\left(\frac{1}{45}\left[18\sqrt{10}\cos\left(\frac13\arccos \frac{23\sqrt{10}}{120}
\right)-1\right]\right)^{\!\frac1\gamma}\!$, 
and we have $T_c^{(3)}(\gamma) \sim\!\frac{1}{2\pi}\left(\frac{31}{15}\right)^{\!\frac1\gamma}\!$ when $\gamma\searrow 0$.
 So also $T_c^{(3)}(\gamma) \searrow\frac{1}{2\pi}$ as $\gamma\to\infty$, and it 
blows up to $\infty$ faster than any inverse power of $\gamma$ when $\gamma\searrow 0$.
 {Replacing $T_c^{(3)}(\gamma)\to T_c^{(3)}(g,\gamma)/g$} 
yields (\ref{eq:TcLOWERboundsTHREE}), with (\ref{eq:gamma3x3Kp})--(\ref{eq:gamma3x3Kr}).

\subsection{The lower bound ${T}_c^{(4)}(\gamma)$} 

 The four-angle truncation yields the $4\times 4$ matrix (\ref{eq:Hfour}).
 Its largest eigenvalue $\mathfrak{g}^{(4)}(\gamma)>0$ is obtained by 
setting $\mathfrak{M} = \mathfrak{G}^{(4)}(\gamma)$ in (\ref{eq:4x4spec})--(\ref{eq:DdefM}) 
and choosing the eigenvalue with $j=1$ and $k=1$.
 This yields 
\begin{equation}\label{eq:TcVPapproxQUATRO}
 T_c^{(4)}(\gamma)  =  \tfrac{1}{2\pi}\!\big( \mathfrak{g}^{(4)}(\gamma)\big)^{\frac1\gamma},
\end{equation}
and now {replacing $T_c^{(4)}(\gamma)\to T_c^{(4)}(g,\gamma)/g$} yields (\ref{eq:TcLOWERboundsFOUR}), with the 
symbols $A$, $B$, $C$, $D$, $Z$ given in (\ref{eq:Zdef})--(\ref{eq:Ddef}).

 We remark that it is relatively simple to determine the largest eigenvalue of the $N=4$ problem,
compared to the $N=3$ problem.
 Namely, since in this all-real-roots situation the square roots in Cardano's formula (\ref{eq:4x4spec})
are real positive, it suggests itself to choose $j=1$ and $k=1$. 

\bigskip

Our lower bounds $T_c^{(N)}(\gamma)$ for $N\in\{1,2,3,4\}$ are shown in Fig.~1 in the introduction.

\section{\hspace{-5pt}A rigorous upper bound on~$T_c(\gamma)$}\label{sec:upperB}

 When $T > T_c(\gamma)$, then  the largest eigenvalue $\kappa$ 
of the linear operator 
$\mathfrak{I} -\beta^\gamma\mathfrak{G}(\gamma)$ satisfies $\kappa= 1 -\beta^\gamma\mathfrak{g}(\gamma) >0$.
 In that case $Q_\gamma(\Xi)\geq 0$ on all of $\ell^2(\NN_0)$, with ``$=0$'' if and only if $\Xi=0$.

 Now recall that precisely at $T = T_c(\gamma)$, when $\kappa=0$ so that $(2\pi T_c)^\gamma = \mathfrak{g}$,
the pertinent eigenvalue problem (\ref{eq:ELinXIlin}) for the minimizing mode $\Xi^{\mbox{\tiny{opt}}}$ reads
\begin{equation}\label{eq:EVeqFORtauC}
\big(\mathfrak{g} \mathfrak{I} -\mathfrak{G}\big) \Xi^{\mbox{\tiny{opt}}} = 0,
\end{equation}
which is equivalent to 
\begin{equation}\label{eq:EVfixPTeqFORtauC}
\mathfrak{C}\big(\mathfrak{g}\big)\Xi^{\mbox{\tiny{opt}}} = \Xi^{\mbox{\tiny{opt}}}
\end{equation}
with 
\begin{equation}
\mathfrak{C}\big(\mathfrak{g}\big) := 
\big(\mathfrak{g} \mathfrak{I} +\mathfrak{G}_1\big)^{-1} \big(\mathfrak{G}_2+ \mathfrak{G}_3\big);
\end{equation}
i.e., (\ref{eq:BS}), restated here for the convenience of the reader.
 In the last part of the proof of Theorem~1, we showed that not only is
$1\in \sigma\big(\mathfrak{C}\big(\mathfrak{g}\big)\big)$, 
but 1 is the spectral radius of $\mathfrak{C}\big(\mathfrak{g}\big)$.
 This showed that the non-trivial solution of (\ref{eq:EVfixPTeqFORtauC}), $\Xi^{\mbox{\tiny{opt}}}$, is in the positive
cone $\ell^2_{\geq 0}(\NN_0)$ (after at most choosing the overall sign).

 These two observations together imply that for $T>T_c$, the compact operator 
$\mathfrak{C}\big((2\pi T)^\gamma\big)$ has spectral radius $\rho\big(\mathfrak{C}\big((2\pi T)^\gamma\big)\big) <1$, and
since it is leaving $\ell^2_{\geq 0}(\NN_0)$, invariant, this now means that when $T>T_c$ then 
$\mathfrak{C}\big((2\pi T)^\gamma\big)$ is a contraction mapping on $\ell^2_{\geq 0}$, with $\Xi=0$ as the only fixed point.
 Put differently,
\begin{equation}\label{eq:LIPestim}
\big\| \mathfrak{C}\big((2\pi T)^\gamma\big)(\Xi - \Xi')\big\| \leq L(\gamma,T) \big\|\Xi - \Xi'\big\|,
\end{equation}
with $L(\gamma,T):= \rho\big(\mathfrak{C}\big((2\pi T)^\gamma\big)\big) <1$
the \textsl{Lipschitz constant} for the linear map $\mathfrak{C}\big((2\pi T)^\gamma\big):\ell^2_{\geq 0}\to\ell^2_{\geq 0}$. 

 We next construct an upper bound $T_c^*$ on $T_c$ , given $\gamma>0$, by showing that 
$\rho\big(\mathfrak{C}\big((2\pi T_c^*)^\gamma\big)\big) < 1$. 
 We accomplish this by invoking another well-known lemma, here already taylored to our needs.

\smallskip
\noindent
{\bf Lemma~8}: \textsl{Let $\varkappa>0$ be given.  
 Then 
\begin{equation}
\rho\big(\mathfrak{C}(\varkappa) \big)\leq 
\rho\Big(\big(\varkappa \mathfrak{I} +\mathfrak{G}_1\big)^{-1}\Big)
\rho\big( \mathfrak{G}_2 + \mathfrak{G}_3\big).
\end{equation}
}
\smallskip

 Recall that $\big((2\pi T)^\gamma \mathfrak{I} +\mathfrak{G}_1\big)^{-1}$ is a diagonal operator with all positive entries, 
and that $\mathfrak{G}_1$ has non-negative entries including 0. 
 Hence, 
\begin{equation}\label{eq:diagEST}
\hspace{-1truecm}
\rho\Big(\!\Big((2\pi T)^\gamma \mathfrak{I} +\mathfrak{G}_1\Big)^{-1}\Big) = 
\max_{n\geq 0} \frac{1}{
(2\pi T)^\gamma + \frac{1}{2 n + 1} {\textstyle\sum\limits_{k=1}^{n}} \frac{2}{ k^\gamma}}
= \frac{1}{(2\pi T)^\gamma}.
\end{equation}
 Since the operator $ \mathfrak{G}_2 + \mathfrak{G}_3$ is independent of $T$, we now arrive at the following conclusion.

\smallskip
\noindent
{\bf Proposition~4}: \textsl{Let $\gamma>0$ be given. Suppose 
\begin{equation}\label{eq:PROPhypo}
T \geq  \tfrac{1}{2\pi}\Big(\rho\big( \mathfrak{G}_2 + \mathfrak{G}_3\big)\Big)^{\frac1\gamma}.
\end{equation}
 Then  $T \geq T_c$.}
\smallskip

\noindent
\textsl{Proof}: By Lemma~8 and by (\ref{eq:diagEST}), and by the hypothesis (\ref{eq:PROPhypo}) of the Proposition, we have that
\begin{equation}\label{eq:specRADchainEST}
\rho\big(\mathfrak{C}\big((2\pi T)^\gamma\big)\big) \leq \frac{1}{(2\pi T)^\gamma} 
\rho\big( \mathfrak{G}_2 + \mathfrak{G}_3\big) 
\leq 1 
= \rho\big(\mathfrak{C}\big(\mathfrak{g}\big)\big) .
\end{equation}
Since $T\mapsto \rho\big(\mathfrak{C}\big((2\pi T)^\gamma\big) \big)$ is monotonically decreasing, the claim of the Proposition follows.
\hfill {\textbf{Q.E.D.}}

\smallskip
\noindent
{\bf Corollary~2}:
 \textsl{Proposition~4 implies that}
\begin{equation}\label{eq:TcBOUNDup}
 T_c(\gamma) \leq  \tfrac{1}{2\pi}\Big(\rho\big( \mathfrak{G}_2 + \mathfrak{G}_3\big)\!\Big)^{\frac1\gamma}.
\end{equation}

We next recall another lemma.
\smallskip

\noindent
{\bf Lemma~9}: \textsl{Let $\gamma>0$ be given.  
 Then 
\begin{equation}
\rho\big( \mathfrak{G}_2 + \mathfrak{G}_3\big)
\leq 
\rho\big( \mathfrak{G}_2\big) + \rho\big(\mathfrak{G}_3\big).
\end{equation}
}

We next estimate these two spectral radii from above. 

\smallskip
\noindent
{\bf Proposition~5}: \textsl{Let $\gamma>0$ be given.  Then for any $\eps\in(0,\min\{2\gamma,1\})$,}
\begin{equation}\label{eq:PROP3}
\rho\big( \mathfrak{G}_2(\gamma)\big) \leq  \big( (2^{1+\eps}-1) \zeta(1+2\gamma-\eps) \zeta(1+\eps) \big)^{\frac12} .
\end{equation}

\noindent
\textsl{Proof}: By the symmetry of $\mathfrak{G}_2$, and its vanishing diagonal elements, 
 \begin{align}\label{eq:SYMofK2}
\hspace{-0.8truecm}
& \sum\limits_{n\geq 0}\sum\limits_{m\geq 0}     
\xi_n  \biggl[\frac{1}{\sqrt{2 n + 1}}\,\frac{1 - \delta_{n,m}}{ |n-m|^\gamma}\, \frac{1}{\sqrt{2 m + 1}}\biggr] \xi_m  = 
\\ \notag &\, 
2\sum_{n\geq 0} \sum_{m>n} 
\xi_n  \biggl[\frac{1}{\sqrt{2 n + 1}}\,\frac{1}{ (m-n)^\gamma}\, \frac{1}{\sqrt{2 m + 1}}\biggr] \xi_m .
\end{align}
 By the Cauchy--Schwarz inequality,  
 \begin{align}\label{eq:CSforK2}
\hspace{-1truecm}
& \sum_{n\geq 0} \sum_{m>n} 
\xi_n  \biggl[\frac{1}{\sqrt{2 n + 1}}\,\frac{1}{ (m-n)^\gamma}\, \frac{1}{\sqrt{2 m + 1}}\biggr] \xi_m \leq \\
& \notag
\biggl(\sum_{k\geq 0} \xi_k^2 \biggr)^{\frac12}\!\!
\biggl(\sum_{n\geq 0} \frac{1}{{2 n + 1}}
\biggl[\sum_{m>n} \frac{1}{ (m-n)^\gamma}\, \frac{1}{\sqrt{2 m + 1}}\xi_m \biggr]^2\biggr)^{\frac12}\!,
\end{align}
and $\sum_{k\geq 0} \xi_k^2 = \| \Xi\|^2_{\ell^2(\NN_0)}$.
 Next, with $\eps\in(0,\min\{2\gamma,1\})$ we rewrite 
\begin{equation}\label{eq:trick17}
\frac{1}{\sqrt{2 m + 1}}\xi_m = \frac{1}{(2 m + 1)^{\frac12-\frac\eps2}}\,\frac{1}{(2 m + 1)^{\frac\eps2}}\xi_m,
\end{equation}
then use the Cauchy--Schwarz inequality again, and obtain
 \begin{align}\label{eq:CSforK2again}
\hspace{-1truecm}
&  \sum_{m>n} \frac{1}{ (m-n)^\gamma}\, \frac{1}{\sqrt{2 m + 1}} \xi_m 
\leq
\\ \notag &\, 
\biggl(\sum_{m>n}\frac{1}{(m-n)^{2\gamma}}\, \frac{1}{(2 m + 1)^{1-\eps}} \biggr)^{\frac12}\!\!
\biggl(\sum_{m>n} \frac{1}{(2 m + 1)^{\eps}} \xi_m^2 \biggr)^{\frac12}\!\!.
\end{align}
 Now
\begin{equation}\label{eq:monoESTIM}
\sum_{m>n} \frac{1}{(2 m + 1)^{\eps}} \xi_m^2 \leq  \frac{1}{(2 n + 1)^{\eps}} \sum_{m>n}\xi_m^2,
\end{equation}
and $\sum_{m>n} \xi_m^2 \leq \| \Xi\|^2_{\ell^2(\NN_0)}$.
 Moreover, since $2m+2n+1> 2m$ and $\eps\in(0,\min\{2\gamma,1\})$, hence $\eps<1$, we estimate
 \begin{align}
\hspace{-1truecm}
\sum_{m>n}\frac{1}{(m-n)^{2\gamma}}\, \frac{1}{(2 m + 1)^{1-\eps}} & = 
\sum_{m>0}\frac{1}{ m^{2\gamma}}\, \frac{1}{(2 m + 2n+ 1)^{1-\eps}} \\ & \leq 
 \frac{1}{2^{1-\eps}} \sum_{m>0}\frac{1}{ m^{1+2\gamma-\eps}},
\label{eq:ZETAforK2}
\end{align}
and since the  hypothesis $\eps\in(0,\min\{2\gamma,1\})$ implies $\eps<2\gamma$, 
the sum at r.h.s.(\ref{eq:ZETAforK2})  exists and equals $\zeta(1+2\gamma-\eps)$.
 Finally, with (\ref{eq:CSforK2})--(\ref{eq:ZETAforK2}), r.h.s.(\ref{eq:SYMofK2}) is estimated by 
\begin{align}\label{eq:SYMofK2estim}
\hspace{-0.8truecm}
 \sum_{n\geq 0} \sum_{m>n} 
\xi_n  \biggl[\frac{1}{\sqrt{2 n + 1}}\,\frac{1}{  (m-n)^\gamma}\, \frac{1}{\sqrt{2 m + 1}}\biggr] \xi_m 
& \leq \\ \notag \, 
 \| \Xi\|^2_{\ell^2(\NN_0)} \biggl(\frac{1}{2^{1-\eps}} \zeta(1+2\gamma-\eps)
\sum_{n\geq 0}\frac{1}{(2n+1)^{1+\eps}}\biggr)^{\frac12}, &
\end{align}
and the remaining sum under the square root equals $(1-\frac{1}{2^{1+\eps}})\zeta(1+\eps)$.
 Multiplying r.h.s.(\ref{eq:SYMofK2estim}) by $2$ and dividing by $ \| \Xi\|^2_{\ell^2(\NN_0)}$ yields r.h.s.(\ref{eq:PROP3}).
\hfill {\textbf{Q.E.D.}}
\smallskip

\noindent
{\bf Proposition~6}: \textsl{Let $\gamma>0$ be given.  Then for any $\eps\in(0,\min\{2\gamma,1\})$,}
\begin{equation}\label{eq:PROP4}
\rho\big( \mathfrak{G}_3(\gamma)\big) \leq 1+ \big((2^{1+\eps}-1) \zeta(1+2\gamma-\eps)\zeta(1+\eps) \big)^{\frac12}. 
\end{equation}
\smallskip

\noindent
\textsl{Proof}: By the symmetry of $\mathfrak{G}_3$, and its non-vanishing diagonal elements, we have 
 \begin{align}\label{eq:SYMofK3}
\hspace{-0.8truecm}
& \sum\limits_{n\geq 0} \sum\limits_{m\geq 0}     
\xi_n  \biggl[\frac{1}{\sqrt{2 n + 1}}\, \frac{1}{ (n+m+1)^\gamma}\, \frac{1}{\sqrt{2 m + 1}}\biggr] \xi_m = 
\\ \notag &\,
\sum_{n\geq 0} 
 \biggl[\frac{1}{ (2n+1)^{1+\gamma}}\, \biggr] \xi_n^2 
\\ \notag
&\ +
2\sum_{n\geq 0} \sum_{m>n} 
\xi_n  \biggl[\frac{1}{\sqrt{2 n + 1}}\, \frac{1}{ (n+m+1)^\gamma}\, \frac{1}{\sqrt{2 m + 1}}\biggr] \xi_m .
\end{align}
For the single sum at r.h.s.(\ref{eq:SYMofK3}) we estimate
\begin{equation}\label{eq:K3singleSUMestim}
\sum_{n\geq 0} 
 \biggl[\frac{1}{(2n+1)^{1+\gamma}}\, \biggr] \xi_n^2 \leq
\|\Xi\|^2_{\ell^2(\NN_0)}.
\end{equation}
 For the double sum at r.h.s.(\ref{eq:SYMofK3}) the Cauchy--Schwarz inequality gives
 \begin{align}\label{eq:CSforK3}
&
\hspace{-1truecm} \sum_{n\geq 0} \sum_{m>n} 
\xi_n  \biggl[\frac{1}{\sqrt{2 n + 1}}\, \frac{1}{ (n+m+1)^\gamma}\, \frac{1}{\sqrt{2 m + 1}}\biggr] \xi_m 
 \leq \\  \notag
 &\hspace{-1truecm} \biggl(\sum_{n\geq 0} \xi_n^2 \biggr)^{\frac12}\!\!
\biggl(\sum_{n\geq 0} \frac{1}{{2 n + 1}}
\biggl[\sum_{m>n} \frac{1}{  (n+m+1)^\gamma}\, \frac{1}{\sqrt{2 m + 1}}\xi_m \biggr]^2\biggr)^{\frac12},
\end{align}
and one more time we note that $\sum_{n\geq 0} \xi_m^2 = \| \Xi\|^2_{\ell^2(\NN_0)}$.
 Using next (\ref{eq:trick17}) again, and then the Cauchy--Schwarz inequality again, we obtain
 \begin{align}\label{eq:CSforK3again}
&\hspace{-.5truecm}  \sum_{m>n} \frac{1}{ (n+m+1)^\gamma}\, \frac{1}{\sqrt{2 m + 1}} \xi_m 
\leq\\ 
\notag &
\hspace{-.5truecm}\, 
\biggl(\sum_{m>n}\frac{1}{(n+m+1)^{2\gamma}}\, \frac{1}{(2 m + 1)^{1-\eps}} \biggr)^{\frac12}\!\!
\biggl(\sum_{m>n} \frac{1}{(2 m + 1)^\eps} \xi_m^2 \biggr)^{\frac12}\!\!.
\end{align}
 The last sum in (\ref{eq:CSforK3again}) is again estimated by (\ref{eq:monoESTIM}), followed by the estimate
 $\sum_{m>n} \xi_m^2 \leq \| \Xi\|^2_{\ell^2(\NN_0)}$.
 The first factor is estimated by
 \begin{align}
\hspace{-.5truecm}\sum_{m>n}\frac{1}{(n+m+1)^{2\gamma}}\, \frac{1}{(2 m + 1)^{1-\eps}}  = &  \\
\hspace{-.5truecm}\sum_{m>0}\frac{1}{(m+2n+1)^{2\gamma}}\, \frac{1}{(2 m + 2n+ 1)^{1-\eps}}  \leq &\\ 
\hspace{-.5truecm} \tfrac{1}{2^{1-\eps}}
\sum_{m>0}\frac{1}{ m^{1+2\gamma-\eps}}  =& \tfrac{1}{2^{1-\eps}}
\zeta(1+2\gamma-\eps), \label{eq:ZETAforK3}
\end{align}
where we used that $2n+1>0$.
 Finally, with (\ref{eq:K3singleSUMestim})--(\ref{eq:ZETAforK3}), r.h.s.(\ref{eq:SYMofK3}) is estimated by 
\begin{align}\label{eq:SYMofK3estim}
\hspace{-1truecm}
 \sum_{n\geq 0} \sum_{m>n} 
\xi_n  \biggl[\frac{1}{\sqrt{2 n + 1}}\,\frac{1}{ (n+m+1)^\gamma}\, \frac{1}{\sqrt{2 m + 1}}\biggr] \xi_m 
& \leq \\ \notag \, 
\hspace{-1truecm} \| \Xi\|^2_{\ell^2(\NN_0)}
\biggl(1 + 2 \biggl(\frac{1}{2^{1-\eps}} \zeta(1+2\gamma-\eps)\sum_{n\geq 0}\frac{1}{(2n+1)^{1+\eps}}\biggr)^{\frac12}\biggr), 
&
\end{align}
and the remaining sum under the square root equals $(1-\frac{1}{2^{1+\eps}})\zeta(1+\eps)$.
 Dividing r.h.s.(\ref{eq:SYMofK3estim}) by $\| \Xi\|^2_{\ell^2(\NN_0)}$ yields r.h.s.(\ref{eq:PROP4}).
\hfill {\textbf{Q.E.D.}}
\smallskip

{From Corollary~2 in concert with Propositions~5 and 6, we now obtain an explicit upper bound on $T_c(\gamma)$.
\smallskip

\noindent
{\bf Theorem~3}: \textsl{Let $\gamma>0$ given.
Then  ${T}_c(\gamma) \leq {T}_c^*(\gamma)$, with
\begin{equation}\label{eq:TcUPPERboundOmegaNULL}
\hspace{-1truecm}
{T}_c^*(\gamma) 
= \frac{1}{2\pi}
\!\left[1+ 
2\big((2^{1+\epsilon(\gamma)}\!-1)\zeta(1\!+\!\epsilon(\gamma)) \zeta(1+2\gamma-\epsilon(\gamma)) \big)^{\frac12}\right]^{\frac1\gamma}\!\!\!,\!\!\!
\end{equation}
where $\epsilon(\gamma) := \min\{\gamma,0.65\}$.
 R.h.s.(\ref{eq:TcUPPERboundOmegaNULL})$\sim \frac{1}{2\pi}\left(\frac2\gamma\right)^{\!\frac1\gamma}$ when $\gamma\searrow 0$,
and $\searrow \frac{1}{2\pi}$ as $\gamma\to\infty$.
}
\smallskip

\noindent
\textsl{Proof}: Clearly, $\epsilon(\gamma) := \min\{\gamma,0.65\}$ 
satisfies the hypothesis on $\eps$ stated in Propositions~5 and 6, for all $\gamma>0$.
 The upper bound on ${T}_c(\gamma)$ thus follows immediately from Corollary~2, Lemma~9, and 
Propositions~5 and~6.
 The asymptotics is easily established.
\hfill {\textbf{Q.E.D.}}
\smallskip

The graph of $T_c^*(\gamma)$ is shown in Fig.~1, together with several lower bounds $T_c^{(N)}(\gamma)$.}

We remark that the particular value of $0.65$ for $\eps$ is suggested by replacing $\epsilon(\gamma)$
with $\eps$ at r.h.s.(\ref{eq:TcUPPERboundOmegaNULL}), then setting $\gamma=2$, and then plotting
the resulting expression vs. $\eps$.
 One then notices that a minimum occurs for $\eps\approx 0.65$.
 For small enough $\gamma$ the value $\eps= 0.65$ violates the hypotheses on $\eps$, though.
 Replacing $\eps$ with $\epsilon(\gamma)$ takes care of this.

 {Replacing $T_c^*(\gamma)\to T_c^*(g,\gamma)/g$ in (\ref{eq:TcUPPERboundOmegaNULL})}
 yields (\ref{eq:TcUPPERboundOmNULL}).
\newpage

\section{Summary and Outlook}

 \subsection{Summary}
 For a version of the Eliashberg theory known as the $\gamma$ model \cite{MoonChubukov} we have conducted a rigorous 
linear stability analysis of the normal state. 
 The $\gamma$ model features two parameters, $g>0$ and $\gamma>0$.
 We established that for each $\gamma>0$ and $g>0$ there exists a critical temperature $T_c(g,\gamma)>\frac{g}{2\pi}$ 
such that the normal state is linearly stable against any small perturbations for $T>T_c$ and unstable against 
superconducting perturbations for $T<T_c$.
 
 We formulated a variational principle for the critical temperature $T_c(g,\gamma)$ in terms of the largest positive 
eigenvalue of an explicitly given compact operator $\mathfrak{G}(\gamma)$ on the Hilbert space of square-summable sequences.
 By restricting the variational principle to an increasing sequence of $N$-dimensional subspaces, 
our variational principle yields explicit lower bounds $T_c^{(N)}(g,\gamma)$ on $T_c(g,\gamma)$ that can be expressed
in closed form when $N\in\{1,2,3,4\}$.
 If demanded, the accuracy can be arbitrarily increased by increasing $N$ beyond $N=4$, thanks to Lemma~6, but the lower bounds 
then have to be approximated by using one of the many available numerical methods to approximately
compute the largest eigenvalue of $\mathfrak{G}^{(N)}$ with acceptable accuracy.

{For $\gamma=2$, our rigorous bound $T_c^{(3)}(g,2)\approx 0.18204 g$ agrees to three significant digits with the 
{approximate} numerical $T_c(g,2)$ value reported in \cite{AllenDynes}, computed using a 64 Matsubara frequency 
truncation of the Eliashberg gap equation.
 Thus, empirically the three-angle truncation seems already remarkably accurate when a three-digit precision suffices 
--- if one assumes that the computations with $N=64$ Matsubara frequencies in \cite{AllenDynes} have converged to 
at least three significant digits.
 However, already the explicit formula obtained for the
truncation with $N=4$ Matsubara frequencies shows a value of $T_c^{(4)}(g,2)\approx 0.1825...g$, which establishes
that the $N=64$ computations of Allen and Dynes \cite{AllenDynes}, if they have converged to more than
two significant digits, correctly produced only the first two of these.
 This is an interesting lesson about computer-empirical evidence.

Nowadays one can easily compute with many more Matsubara frequencies.
 Increasing $N$ beyond 200, the first 10 significant digits of the sequence of lower bounds $T_c^{(N)}(g,\gamma)$ have
stabilized, yielding $T_c(g,2) = 0.1827262477...g$ independently with Maple and with Mathematica numerical algorithms; note 
that none of the displayed digits has been rounded.
 Needless to say that such a numerical precision is beyond the possibility of rigorous upper and lower estimates.

 Reformulating the linearized Eliashberg gap equation as a fixed point problem, we also obtained a rigorous upper bound
on $T_c(g,\gamma)$ expressed in closed form. 
While this bound is more than twice as large than the numerical $T_c$ value at $\gamma=2$, it becomes asymptotically exact
with large $\gamma$. 
 Indeed, $T_c^*(g,\gamma)$ does converge to $\frac{g}{2\pi}$ from above when $\gamma\to \infty$. 
 Also all lower bounds on $T_c(g,\gamma)$ converge to $\frac{g}{2\pi}$  when $\gamma\to\infty$.
 Therefore, $T_c(g,\infty) = \frac{g}{2\pi}$. 

The implied tightness of the lower and upper bounds for when $\gamma \gg 1$ implies that for larger $\gamma$
values the lower bounds converge more and more rapidly to $T_c(g,\gamma)$.
 This may not be a cause for much celebration, though, as the empirically useful $\gamma$ values so far are all $\leq 2$
(recall that formally $\gamma=2$ has physical significance in the sense that for $\gamma=2$ the 
$\gamma$ model is obtained through a rescaled limit $\lambda\to\infty$ of the {Einstein phonon} model).
Empirical realms aside, the large $\gamma$ regime is of independent theoretical interest, see \cite{YuzKieAltPRB}.}

 For small $\gamma$ the rate of convergence of the lower bounds seems to slow down dramatically. 
 We note that all lower bounds blow up to $\infty$ when $\gamma\searrow 0$ like $C_N^{\frac1\gamma}$, with $C_N$ itself
increasing with $N$ beyond any bound.
 Our upper bound on $T_c(g,\gamma)$ sets the upper bound $C \left(\frac2\gamma\right)^{\!\frac1\gamma}$ 
on how fast $T_c(g,\gamma)$ can diverge to $\infty$ when $\gamma\searrow 0$, which is compatible with the estimated blow-up
behavior {given in}   \cite{WAAYC}.
 The asymptotic constant $C$ in our upper bound is presumably not sharp, though.
\vspace{-10pt}

 \subsection{Outlook}\vspace{-5pt}

 In our follow-up paper, i.e. part II of our series on $T_c$ in the Eliashberg theory, we will address the standard
version of Eliashberg theory in which the effective electron-electron interactions are mediated by generally dispersive phonons.
 Since the $\gamma$ model at $\gamma=2$ captures the asymptotic $\lambda\to\infty$ behavior of the standard version, the 
results obtained in the present paper will serve us as important asymptotic input for controlling the large $\lambda$ regime. 
 We emphasize that our control of the standard version of Eliashberg theory will be not merely in terms of some asymptotic 
expansion around $\lambda=\infty$.
 It will also supply rigorous convergence results valid for all $\lambda >0$. 

 Moreover, and more subtly, to control the standard version of Eliashberg theory we will need some results about the $\gamma$ 
model for values of $\gamma$ other than $\gamma=2$, in particular for $\gamma=4$. 
 This is not a value distinguished by empirical physics which, so far, has seen 
applications only for  certain values of $\gamma<2$~\cite{MoonChubukov,ChubukovETal}.
 
 Subsequently, in part III of our series of papers on $T_c$ in Eliashberg theory, we will address the dispersionless limit
of the standard version of Eliashberg theory.
 The dispersionless limit features Einstein phonons of frequency $\Omega$, yet is also known as the Holstein model.
  Several results for the $\gamma$ model will enter again.
 Since the normalized frequency measure $P(d\omega)$ of the standard version degenerates into $\delta(\omega-\Omega)d\omega$ 
for the Holstein model, several general expressions become explicitly evaluable, yielding additional insights into the 
materials governed by optical phonons.

 \darkred{We close with the remark that the mathematical treatment of the $\gamma$ model presented in this paper can presumably 
be directly adapted to study the extended $\gamma$ model of \cite{WZAC}.}
 
\bigskip

{\bf Acknowledgement:} We thank Artem Abanov, Andrey Chubukov, and Yiming Wu for their comments on the preprint version of our work, and
for making their numerical $T_c(\gamma)$ data available. 
 Thanks go also to Steven Kivelson for his comments. 
Finally we thank the three referees for their constructive criticisms. 

\vfill
DATA AVAILABILITY STATEMENT: \magenta{The numerical data that we have produced for this paper will be made available upon reasonable request.}
\darkred{For the numerical data of \cite{WAC} please contact the authors of \cite{WAC}.}
\bigskip

CONFLICT OF INTEREST STATEMENT: The authors declare that they have no conflict of interest.
\newpage

\begin{appendices}
\section{Compactness of $\mathfrak{G}$}\label{appendixA}

In this appendix we establish that the three operators $\mathfrak{G}_j$, $j\in\{1,2,3\}$, are Hilbert--Schmidt operators, 
hence compact.
\smallskip

\noindent
{\bf Proposition~A}: \textsl{For any $\gamma>0$, $\mathfrak{G}_1\in \ell^2(\NN_0)\times\ell^2(\NN_0)$.}
\smallskip

\noindent
\textsl{Proof}: The operator $\mathfrak{G}_1$ is diagonal, with elements
 \begin{align}\label{eq:Kop1nm}
(\mathfrak{G}_1)_{n,m} = 
 \biggl[\frac{1}{2 n + 1} {\sum\limits_{k=1}^{n}} \frac{2}{ k^\gamma}\biggr] \delta^{}_{n,m}.
\end{align}
 Squaring (\ref{eq:Kop1nm}) and summing $n$ and $m$ over $\NN_0$ yields
 \begin{align}\label{eq:Kop1nmSQ}
{\sum\limits_{n\in\NN_0}\!\sum\limits_{m\in\NN_0}  
(\mathfrak{G}_1)_{n,m}^2 = 
\sum\limits_{n\in\NN_0} \frac{1}{(2 n + 1)^2} \biggl[ {\sum\limits_{k=1}^{n}} \frac{2}{k^\gamma}\biggr]^2}.
\end{align}
{A Riemann sum approximation now yields
\begin{equation}
\sum_{k=1}^{n}{\frac{1}{k^\gamma}} 
\leq 
1 + \left\{ \genfrac{}{}{0pt}{0}{\frac{1}{1-\gamma} \left(n^{1-\gamma}-1\right);\ \gamma\neq 1}
{\qquad\quad \ln n\ \qquad;\ \gamma =1}\right. .
\end{equation}
}
 Hence $\sum\limits_n\!\sum\limits_m  (\mathfrak{G}_1)_{n,m}^2 < \infty$.
\hfill {\textbf{Q.E.D.}}
\smallskip

\noindent
{\bf Proposition~B}: \textsl{For any $\gamma>0$, $\mathfrak{G}_2\in \ell^2(\NN_0)\times\ell^2(\NN_0)$.}
\smallskip

\noindent
\textsl{Proof}: 
The operator $\mathfrak{G}_2$ has elements
 \begin{align}\label{eq:Kop2nm}
(\mathfrak{G}_2)_{n,m} = 
\frac{1}{\sqrt{2 n + 1}}\,\frac{1-\delta_{n,m}}{|n-m|^\gamma}\,\frac{1}{\sqrt{2 m+1}}
\end{align}
 Squaring (\ref{eq:Kop2nm}), then summing $n$ and $m$ over $\NN_0$, 
and using the symmetry $(\mathfrak{G}_2)_{n,m} = (\mathfrak{G}_2)_{m,n}$, we obtain
 \begin{align}\label{eq:Kop2nmSQa}
\hspace{-1truecm}
\sum\limits_{n\in\NN_0}\!\sum\limits_{m\in\NN_0}  
(\mathfrak{G}_2)_{n,m}^2 
& = 
\sum\limits_{n\in\NN_0}\!\sum\limits_{m\in\NN_0}  
 \frac{1}{{2 n + 1}}\biggl[\frac{1-\delta_{n,m}}{|n-m|^\gamma}\biggr]^2\,\frac{1}{{2 m+1}} \\ 
\hspace{-.5truecm} \label{eq:Kop2nmSQb}
& =
2 \sum\limits_{n\geq0}\sum\limits_{m>n} \frac{1}{{2 n + 1}}\,\frac{1}{(m-n)^{2\gamma}}\,\frac{1}{{2 m+1}} .
\end{align}
 By relabeling,
 \begin{align}\label{eq:Kop2nmSQrelabel}
\hspace{-1truecm}
\sum\limits_{m>n} \frac{1}{(m-n)^{2\gamma}}\,\frac{1}{{2 m+1}} = \sum\limits_{m>0} \frac{1}{m^{2\gamma}}\,\frac{1}{{2 m+ 2n+1}}. 
\end{align}

 Now suppose first that $\gamma >\frac12$.
 We simply estimate $2m+2n+1>2n+1$ for $m>0$. 
 The sum over $m$ in (\ref{eq:Kop2nmSQrelabel}) then equals $\zeta(2\gamma)$, multiplied by $\frac{1}{2n+1}$.
 The resulting sum over $n$ at r.h.s.(\ref{eq:Kop2nmSQb}) equals $\frac18\pi^2$.

 Suppose next that $\gamma \leq\frac12$.
 In this case we use the estimate
 \begin{align}\label{eq:Kop2nmSQtrick}
\hspace{-.5truecm}
\frac{1}{(2 m+ 2n+1)} =&\, 
\frac{1}{(2 m+ 2n+1)^{1-\gamma}(2 m+ 2n+1)^{\gamma}} \\ \leq &\,
\frac{1}{(2 m )^{1-\gamma}(2n+1)^{\gamma}} .
\end{align}
 And so
\begin{align}
\hspace{-1truecm}
 \sum\limits_{n\geq0}\sum\limits_{m>n} \frac{1}{{2 n + 1}}\,\frac{1}{(m-n)^{2\gamma}}\,\frac{1}{{2 m+1}} 
\leq & \\ \label{eq:Kop2nmSQbound}
\sum\limits_{n\geq 0}  \frac{1}{(2 n + 1)^{1+\gamma}}\,\sum\limits_{m >n}  \frac{1}{m^{2\gamma}}\,\frac{1}{(2 m)^{1-\gamma}} 
\leq & \\  \notag
\frac{1}{2^{1-\gamma}}\sum\limits_{n\geq 0}  \frac{1}{(2 n + 1)^{1+\gamma}}\sum\limits_{m\geq 1}\frac{1}{m^{1+\gamma}} .&
\end{align}
 At r.h.s.(\ref{eq:Kop2nmSQbound}), 
the sum over $n$ equals $\big(1-\frac{1}{2^{1+\gamma}}\big)\zeta(1+\gamma)$, and the sum over $m$ equals $\zeta(1+\gamma)$.
\hfill {\textbf{Q.E.D.}}
\smallskip

\noindent
{\bf Proposition~C}: \textsl{For any $\gamma>0$, $\mathfrak{G}_3\in \ell^2(\NN_0)\times\ell^2(\NN_0)$.}

\smallskip

\noindent
\textsl{Proof}:
The operator $\mathfrak{G}_3$ has elements
 \begin{align}\label{eq:Kop3nm}
(\mathfrak{G}_3)_{n,m} = 
\frac{1}{\sqrt{2 n + 1}}\,\frac{1}{(n+m+1)^\gamma}\,\frac{1}{\sqrt{2 m+1}}
\end{align}
 Squaring (\ref{eq:Kop3nm}), then summing $n$ and $m$ over $\NN_0$, we obtain
 \begin{align}\label{eq:Kop3nmSQ}
\hspace{-0.8truecm}
\sum\limits_{n\in\NN_0}\!\sum\limits_{m\in\NN_0}  (\mathfrak{G}_3)_{n,m}^2 
 = 
\sum\limits_{n\in\NN_0}\!\sum\limits_{m\in\NN_0} \frac{1}{2 n + 1}\,\frac{1}{(n+m+1)^{2\gamma}}\,\frac{1}{2 m+1} .
\end{align}
 Next we note that {for any $\gamma>0$}
  \begin{align}\label{eq:Kop3nmSQtrick}
\hspace{-1truecm}
\frac{1}{(n+m+1)^{2\gamma}}
\! = &\,
\frac{1}{(n+m+1)^\gamma}\, \frac{1}{(n+m+1)^\gamma} \\ 
\leq &\, 
\frac{1}{(n+1)^\gamma}\, \frac{1}{(m+1)^\gamma} .
\end{align}
 And so, 
 \begin{align}\label{eq:Kop3nmSQbound}
\sum\limits_{n\in\NN_0}\!\sum\limits_{m\in\NN_0}  
(\mathfrak{G}_3)_{n,m}^2 
 \leq
\biggl[\sum\limits_n \frac{1}{(n + 1)^{1+\gamma}}\biggr]^2 = \zeta^2(1+\gamma).
\end{align}
\hfill {\textbf{Q.E.D.}}
\smallskip

\section{The matrix invariants}\label{appendixB} 

 We list the evaluation of the matrix invariants that enter our spectral formulas.

\subsection{Trace and determinant for $\mathfrak{G}^{(2)}$}

One easily finds 
\begin{align}\label{eq:trKtwo}
{\rm tr}\,\mathfrak{G}^{(2)}(\gamma)
 &=
 \frac13 \Big(\frac{1}{3^\gamma} +1 \Big)\\
\label{eq:detKtwo}
\det\mathfrak{G}^{(2)}(\gamma) &= 
 \frac13\Big(-\frac{1}{4^\gamma} +\frac{1}{3^\gamma} - \frac{2}{2^\gamma} - 3 \Big).
\end{align}

 We note that ${\rm tr}\,\mathfrak{G}^{(2)}(\gamma)>0$ and
 $\det\mathfrak{G}^{(2)} (\gamma) <0$  for all $\gamma>0$, as readily seen by inspection.

\subsection{${\rm tr}\, \mathfrak{G}^{(3)}$, ${\rm tr\, adj}\, \mathfrak{G}^{(3)}$, and $\det \mathfrak{G}^{(3)}$}

 With the help of Maple, we computed 
\begin{align}
\label{eq:TR3x3K}
{\rm tr}\,\mathfrak{G}^{(3)}(\gamma) & = 
 \frac{1}{15} \Big( \frac{3}{5^\gamma} + \frac{5}{3^\gamma} - \frac{6}{2^\gamma} - 1 \Big),\\
\label{eq:TRadjKthree}
{\rm tr\, adj}\,\mathfrak{G}^{(3)}(\gamma) & =  
 \frac{1}{15} \Big(-\frac{1}{16^\gamma} + \frac{1}{15^\gamma} - \frac{3}{9^\gamma} - \frac{8}{6^\gamma} 
\\ \notag & \qquad\quad
+ \frac{1}{5^\gamma} - \frac{10}{4^\gamma} + \frac{3}{3^\gamma} - \frac{12}{2^\gamma} - 18\Big),\\
\label{eq:detKthree}
\det\mathfrak{G}^{(3)}(\gamma) & = 
 \frac{1}{15} \Big(
-\frac{1}{27^\gamma} 
+ \frac{2}{24^\gamma}
- \frac{1}{20^\gamma}
- \frac{2}{18^\gamma}
\\ \notag & 
\qquad\quad
+ \frac{1}{16^\gamma}
+ \frac{1}{15^\gamma} 
+ \frac{1}{12^\gamma} 
- \frac{2}{10^\gamma}
+ \frac{2}{9^\gamma}
\\ \notag & 
\qquad\quad
+ \frac{4}{8^\gamma}
+ \frac{4}{6^\gamma} 
+ \frac{8}{4^\gamma} 
- \frac{3}{5^\gamma} 
+ \frac{12}{2^\gamma}
+ 5 \Big).
\end{align}

\subsection{${\rm tr}\,\mathfrak{G}^{(4)}$, ${\rm tr}\,\big(\mathfrak{G}^{(4)}\big)^2$, ${\rm tr}\,\big(\mathfrak{G}^{(4)}\big)^3$,
 and $\det \mathfrak{G}^{(4)}$}

With the help of Maple we have computed
\begin{equation}
\label{eq:TR4x4K}
\mathrm{tr}\, \mathfrak{G}^{(4)}(\gamma) = 
\frac{1}{105}\Big(
 \frac{15}{7^\gamma} 
+\frac{21}{5^\gamma} 
+ \frac{5}{3^\gamma} 
-\frac{72}{2^\gamma} 
- 37
\Big),
\end{equation}
\begin{align}
\label{eq:TR4x4Ksquared}
&\hspace{-1truecm}
\mathrm{tr}\, \big(\mathfrak{G}^{(4)}\big)^2(\gamma) 
 = \\ \notag
&\hspace{-1truecm}
\frac{1}{105^2}\Big(
 \frac{225}{49^\gamma}
+\frac{630}{36^\gamma}
+ \frac{1491}{25^\gamma}
- \frac{900}{21^\gamma}
+ \frac{4620}{16^\gamma}
- \frac{900}{14^\gamma}
+ \frac{6300}{12^\gamma}
+\frac{336}{10^\gamma}
+ \frac{9685}{9^\gamma}
\\ \notag
& 
-\frac{900}{7^\gamma}
+ \frac{11880}{6^\gamma}
- \frac{1764}{5^\gamma}
+ \frac{18414}{4^\gamma}
- \frac{3100}{3^\gamma}
+ \frac{20028}{2^\gamma}
+ 28039\Big),
\end{align}
\begin{align}
\label{eq:TR4x4Kcubed}
&\hspace{-1truecm}
\mathrm{tr}\, \big(\mathfrak{G}^{(4)}\big)^3(\gamma) 
 = \\ \notag
&\hspace{-1truecm}
\frac{1}{105^3}\Big(
 \frac{3375}{343^\gamma}
+ \frac{14175}{252^\gamma}
+ \frac{19845}{180^\gamma}
+ \frac{23625}{175^\gamma}
- \frac{20250}{147^\gamma}
+ \frac{9261}{125^\gamma} 
+ \frac{66150}{120^\gamma}
\\ \notag
& 
+ \frac{70875}{112^\gamma}
- \frac{28350}{108^\gamma}
- \frac{20250}{98^\gamma}
+ \frac{141750}{84^\gamma}
+ \frac{46305}{80^\gamma}
+ \frac{7875}{75^\gamma}
\\ \notag
& 
+ \frac{130410}{72^\gamma}
+ \frac{47250}{70^\gamma}
+ \frac{111375}{63^\gamma}
+ \frac{198450}{54^\gamma}
- \frac{102816}{50^\gamma}
- \frac{20250}{49^\gamma}
\\ \notag
& 
+ \frac{200025}{48^\gamma}
+ \frac{138915}{45^\gamma}
+ \frac{109350}{42^\gamma}
+ \frac{330750}{40^\gamma}
- \frac{153090}{36^\gamma}
- \frac{234360}{32^\gamma}
\\ \notag
& 
+ \frac{730170}{30^\gamma}
+ \frac{64125}{28^\gamma}
- \frac{125875}{27^\gamma}
- \frac{213066}{25^\gamma}
+ \frac{179550}{24^\gamma}
+ \frac{81000}{21^\gamma}
\\ \notag
& 
+ \frac{645057}{20^\gamma}
- \frac{557280}{18^\gamma}
+ \frac{901215}{16^\gamma}
+ \frac{330750}{15^\gamma}
+ \frac{81000}{14^\gamma}
+ \frac{1542510}{12^\gamma}
\\ \notag
& 
- \frac{92736}{10^\gamma}
+ \frac{631320}{9^\gamma}
+ \frac{447012}{8^\gamma}
+ \frac{54675}{7^\gamma}
+ \frac{2299410}{6^\gamma}
+ \frac{243432}{5^\gamma}
\\ \notag
& 
+ \frac{310986}{4^\gamma}
+ \frac{1331250}{3^\gamma}
+ \frac{837036}{2^\gamma}
+ 784412
\Big)
\end{align}
and
\begin{align}
\label{eq:DET4x4Kcubed}
&\hspace{-1.truecm}
\mathrm{det}\,\mathfrak{G}^{(4)}(\gamma) = \\ \notag
&\hspace{-1.truecm}
\frac{1}{105^{}}\Big(
\frac{1}{256^\gamma}
- \frac{7}{240^\gamma}
+ \frac{1}{225^\gamma}
+ \frac{2}{216^\gamma}
+ \frac{2}{200^\gamma}
- \frac{1}{189^\gamma}
- \frac{14}{180^\gamma}
+ \frac{2}{168^\gamma}   
\\ \notag   
&
+ \frac{2}{162^\gamma}
- \frac{2}{160^\gamma}
+ \frac{4}{150^\gamma}
+ \frac{2}{144^\gamma}
- \frac{1}{140^\gamma}
- \frac{1}{135^\gamma}
- \frac{2}{126^\gamma}
 \\ \notag
&
- \frac{1}{125^\gamma}
- \frac{2}{120^\gamma}
+ \frac{1}{112^\gamma}
+ \frac{1}{108^\gamma}  
+ \frac{1}{105^\gamma} 
+ \frac{7}{100^\gamma}
- \frac{2}{96^\gamma} 
+ \frac{1}{84^\gamma}
\\ \notag
& 
+ \frac{2}{81^\gamma}
+ \frac{2}{75^\gamma}
- \frac{8}{72^\gamma}
- \frac{2}{70^\gamma} 
+ \frac{2}{63^\gamma}
+ \frac{4}{60^\gamma}
+ \frac{4}{56^\gamma}
+ \frac{4}{54^\gamma}
- \frac{8}{48^\gamma}
\\ \notag
&
- \frac{2}{45^\gamma}
+ \frac{4}{42^\gamma}
- \frac{4}{40^\gamma}
+ \frac{14}{36^\gamma}
- \frac{7}{35^\gamma}
- \frac{28}{32^\gamma}
- \frac{14}{30^\gamma}
+ \frac{8}{28^\gamma} 
+ \frac{2}{27^\gamma}
\\ \notag
&
+ \frac{2}{25^\gamma}
- \frac{28}{24^\gamma}
+ \frac{5}{20^\gamma} 
- \frac{16}{18^\gamma}
- \frac{56}{16^\gamma}
- \frac{2}{15^\gamma} 
+ \frac{28}{14^\gamma}
- \frac{44}{12^\gamma} 
\\ \notag
&
+ \frac{28}{10^\gamma}
- \frac{11}{9^\gamma}
- \frac{32}{8^\gamma}
+  \frac{5}{7^\gamma}
- \frac{38}{6^\gamma}
+ \frac{8}{5^\gamma} 
 - \frac{41}{4^\gamma}
- \frac{13}{3^\gamma}
-  \frac{30}{2^\gamma}
- 7
\Big)
\end{align}

\end{appendices}

\end{document}